\documentclass[sigconf]{acmart}
\usepackage[utf8]{inputenc}
\usepackage[rightcaption]{sidecap}
\sidecaptionvpos{figure}{c}
\usepackage{graphicx}
\usepackage{subcaption}
\usepackage{multirow}
\usepackage{hyperref}
\usepackage{pgfplots}
\usepackage{accsupp}
\usepackage{enumerate}
\usepackage{enumitem}
\usepackage{pifont}
\usepackage[para]{footmisc}
\usepackage[show]{chato-notes}
\usepackage{array}
\usepackage{caption}
\usepackage{nopageno}
\newcolumntype{P}[1]{>{\centering\arraybackslash}p{#1}}

\definecolor{swss}{rgb}{0.57, 0.36, 0.51}

\newcommand{\craig}[1]{\textcolor{black}{#1}}
\newcommand{\iadh}[1]{\textcolor{black}{#1}}
\newcommand{\zm}[1]{\textcolor{black}{#1}}
\newcommand{\siwei}[1]{\textcolor{black}{#1}}

\newcommand{\zms}[1]{\textcolor{black}{#1}} % changes by Zaiqiao
\newcommand{\zmss}[1]{\textcolor{black}{#1}} % Second round changes by Zaiqiao
\newcommand{\sws}[1]{\textcolor{black}{#1}}
\newcommand{\swss}[1]{\textcolor{black}{#1}} % Second round changes by Siwei Liu
\newcommand{\io}[1]{\textcolor{black}{#1}}
\newcommand{\cm}[1]{\textcolor{black}{#1}}

\newcommand{\zmrs}[1]{\textcolor{black}{#1}} % changes by Zaiqiao.
\newcommand{\swrs}[1]{\textcolor{black}{#1}} % changes by Zaiqiao.
\newcommand{\ya}[1]{\textcolor{black}{#1}} % changes by Iadh
\newcommand{\cmrs}[1]{\textcolor{black}{#1}} % changes by Craig.

\usepackage{amsmath, bm}
\newcommand{\mat}[1]{{\bm{#1}}}
\newcommand{\set}[1]{\mathcal{#1}}
\newcommand{\fig}{Figure}
\newcommand{\tab}{Table}
\newcommand{\eq}{Eq. }
\newcommand{\gcn}{\textbf{GCN-P}}
\newcommand{\com}{\textbf{COM-P}}

\usepackage{marginnote}

\author{Zaiqiao Meng}
\authornote{Both authors contributed equally to this work.}
\affiliation{%
  \institution{University of Glasgow}
}
\email{{zaiqiao.meng@gmail.com}}
\author{Siwei Liu}
\authornotemark[1]
\affiliation{%
	\institution{University of Glasgow,}
}
\email{{s.liu.4@research.gla.ac.uk}}
\author{Craig Macdonald}
\affiliation{%
	\institution{University of Glasgow,}
}
\email{{craig.macdonald@glasgow.ac.uk}}
\author{Iadh Ounis}
\affiliation{%
	\institution{University of Glasgow,}
}
\email{{iadh.ounis@glasgow.ac.uk}}

\setcopyright{acmcopyright}
\copyrightyear{2021}
\acmYear{2021}
\pgfplotsset{compat=1.14}
\begin{document}
\title{Graph Neural Pre-training for Enhancing Recommendations using Side Information}
% \title{Graph Neural Pre-training for Recommendation with Side Information}
% \pagestyle{plain} % removes running headers

\begin{abstract}

Leveraging \craig{the} side information associated with entities (i.e.\ users and items) to enhance the performance of recommendation systems has been widely recognized as an important modelling dimension. While many existing approaches focus on the \emph{integration scheme} \craig{to} incorporate entity side information -- by combining the recommendation loss \iadh{function} with an extra \zms{side information-aware} loss -- in this paper, we propose \iadh{instead} a novel \emph{pre-training scheme} for leveraging the side information. \craig{In particular,} we first pre-train a representation model using the side information of \craig{the} entities, and then fine-tune it using an existing general representation-based recommendation model. Specifically, we propose two \zmrs{pre-training} models, \cm{named} \gcn{} and \com{}, by considering the entities and their relations constructed from side information as two different types of graphs respectively, to pre-train entity embeddings. For the \gcn{} model, two single-relational graphs are constructed from \zm{all the users' and items'} side information respectively, to pre-train entity representations by using the Graph Convolutional Networks. For the \com{} model, two multi-relational graphs are constructed to pre-train \iadh{the} entity representations by using the Composition-based Graph Convolutional Networks. 
%With the pre-training procedure of encoding dependencies within users and items from their side information, our models provide the existing recommenders with better generalizability and stability\inote{vague, how is this measured}.
\io{An extensive evaluation} of our pre-training models fine-tuned under four \zmss{general representation-based} recommender models, i.e.\ MF, NCF, NGCF and LightGCN, \io{shows} that effectively pre-training embeddings with both \io{the} user's and item's side information can significantly improve these original models in \craig{terms of} both \io{effectiveness} and stability.
%\inote{stability or robustness?}
\end{abstract}

\maketitle

% ------------ Introduction ------------
\thispagestyle{empty}
\section{Introduction}
\label{sec:Introduction}
% Recommendations
The goal of recommender systems is to assist users in filtering out \iadh{non-relevant} information and selecting \craig{a personalized set of interesting items to} maximize the users' satisfaction. Modern recommendation models achieve this goal by learning representation vectors (i.e.\ embeddings) of the two entities (i.e.\ users and items) that capture the users' \craig{interests} and items' attractiveness~\cite{zhang2017joint,zhang2019deep}, so that the learned embeddings can be used to accurately predict which items a user might choose in the future, e.g., by \siwei{computing} the dot product or a multilayer perceptron (MLP)~\cite{rendle2020neural} \siwei{of the users' and items' embeddings}. Typically, \zms{recommendation models \cm{are collaborative in nature, learning} \io{the} users' \craig{interests} and items' \zm{attractiveness} \iadh{through \io{users'} ratings}}, clicking or other interactive context crossing the two entities\zms{, which we refer to as the \em cross-entity context\em. 
\zmss{For example, the NGCF~\cite{NGCF19} model leverages the high-order context between users and items to enhance a recommender by exploiting the interactive paths \io{across} the two entities.}
However, in real-world data, these cross-entity contextual interactions} are \io{typically} highly sparse~\cite{liu2020heterogeneous}. \craig{Therefore,} a number of side information-aware recommendation models~\cite{ning2012sparse,chen2018collective,vasile2016meta,liu2019recommender,liu2020heterogeneous} \iadh{have been} designed to address this sparsity issue by integrating the rich side information of users and items, such as \iadh{the users'} age groups and \iadh{the items'} \craig{textual} \craig{descriptions}. Such side information about entities can be used to learn the \zms{\em within-entity context \em knowledge to further enhance the recommendation performance.} 
\zmss{For instance, movies with the same features (e.g. same genres and actors) may attract the same users, and such feature relations between movies are a type of knowledge within the side information of movies.}
%\inote{unclear which refers to what? sentence is not easy to read; better to split into 2}
%, demonstrating promising  performance improvements over \craig{models solely} using interactive \iadh{behaviour data}. %the plain ones

% \begin{figure}
% 	\begin{centering}
% 	\includegraphics[width = 120mm]{figs/seed.pdf}
% 	\caption{\label{fig:variance} Box-and-whisker diagrams for the NDCG performances of the MF~\cite{rendle2020neural}, NGCF~\cite{NGCF19} and LightGCN~\cite{he2020lightgcn} models on the Epinions and Foursquare datasets. Large variances can be observed over 50 runs with different random seeds, demonstrating the instabilities of these three models.}
% 	%\zmsc{Update with new results, replace ML\_100K with Epinions.} \sws{Already updated by Siwei}
% 	\end{centering}
% \end{figure}
\begin{figure*}[htp]
  \centering
  \includegraphics[height=8cm]{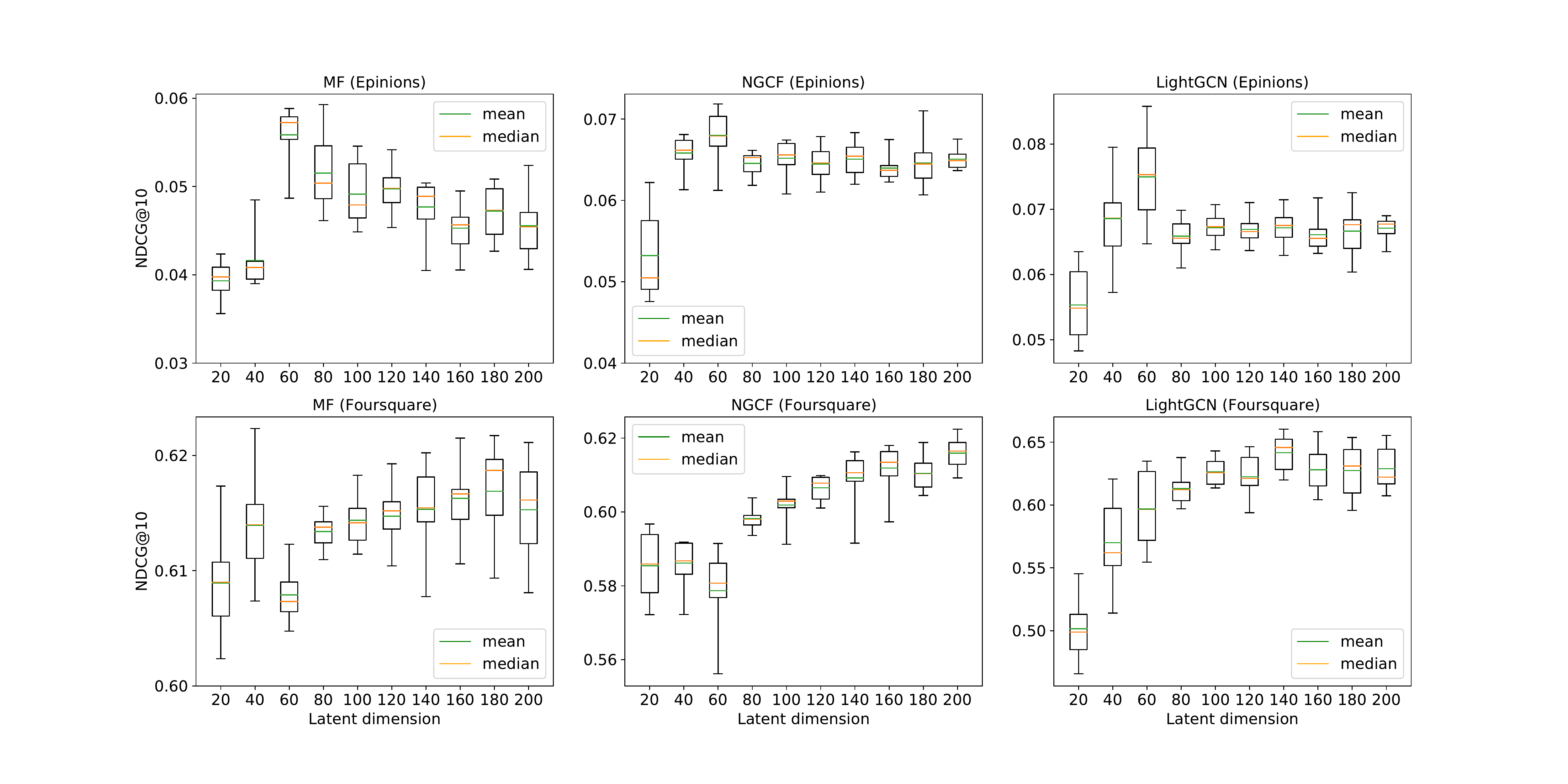}
  \caption{\label{fig:variance}Box-and-whisker diagrams for the NDCG performances of the MF~\cite{rendle2020neural}, NGCF~\cite{NGCF19} and LightGCN~\cite{he2020lightgcn} models on the Epinions and Foursquare datasets. Large variances can be observed over 50 runs with different random seeds, demonstrating the instabilities of these three models.}
\end{figure*}

\thispagestyle{empty}
To leverage \craig{the} side information associated with users and items, many \zm{approaches} have been proposed, \zms{most of which follow the conventional \emph{integration scheme}, \io{which} encodes the side information simultaneously with} the training of user-item interactions~\cite{chen2018collective,park2013hierarchical,liu2019recommender,ning2012sparse}. \zms{They normally optimize} a loss \iadh{function} consisting of two components, i.e. the recommendation loss and \zmrs{one (or even more)} additional \zms{side information}-aware loss ~\cite{liu2019recommender,zhao2017learning,chen2018collective,wang2019kgat,ma2019jointly}.
%, which \iadh{could} \zm{counteract} the effectiveness \iadh{of} each other due to different objectives \iadh{in} the two components\inote{unclear if this counterbalancing is a good or bad thing? ZQ:Could `counteract' make it more like a bad thing?}, \iadh{while ignoring} the independent knowledge of entities for recommendation. 
%\zm{However, such integration scheme may fail in scenarios where the two loss components strongly disagree during training, resulting in reduced effectiveness.}
\zmrs{However, these models require tremendous efforts to tune \zmrs{one (or more)} \ya{hyperparameters} to find a good trade-off solution between the two objectives~\cite{zhao2017learning,liu2019recommender}, and they may fail in scenarios where the two loss components strongly disagree during training, resulting in reduced effectiveness.}
Moreover, \cmrs{while} there \iadh{have been} many  powerful neural network-based recommendation algorithms \iadh{proposed} \iadh{in} recent years~\cite{he2017neural,NGCF19,he2020lightgcn,rendle2020neural}, most of \iadh{these approaches} are unable to give stable recommendation results. \iadh{Indeed, as} we can see from \fig{}~\ref{fig:variance}, with the different random initializations of the \iadh{model's} parameters, \zm{high variances \craig{can be} observed on the performances of \craig{both} the conventional model (i.e. MF~\cite{rendle2020neural}) and the deep neural network-based models (i.e. NGCF~\cite{NGCF19} and LightGCN~\cite{he2020lightgcn})} over different embedding dimensions in both the \sws{Epinions} and Foursquare datasets, \craig{demonstrating the lack of} stability of these models. Therefore, it is crucial to find a \siwei{better} initialization of the entity embeddings and \iadh{the} model parameters, \io{which} allow the stochastic gradient descent algorithm to find a \zms{stable} local optimal recommendation solution.

% Therefore, it is crucial to find an appropriate initialization of the entity embeddings and \iadh{the} model parameters that allow the stochastic gradient descent algorithm to find a good local minimum recommendation solution.

% Appropriate initialization of embeddings is critical

% Recommendation Pretrain (GMF) (VAE)

To address the aforementioned issues, we provide a \em novel \craig{pre-training} scheme \em for leveraging the side information in recommender systems, namely, we \iadh{propose to} pre-train the entity embeddings using the side information, and then fine-tune \iadh{them} using an existing recommendation model. Specifically, we \zms{explore two types of graph-structured data to capture the interdependent relationships among the entities, and propose} two \zmss{pre-training} models based on Graph Neural Networks (GNNs), namely, the \gcn{} model and the \com{} model. The \gcn{} model learns the entity embeddings on the single-relational graphs using Graph Convolutional Networks~\cite{kipf2017semi}, while the \com{} model learns entity embeddings on multi-relational graphs using Composition-based Multi-Relational Graph Convolutional Networks~\cite{vashishth2019composition}. With the \zms{expressive power} of GNNs that recursively \siwei{propagate messages and aggregate features} over neighbours, our pre-training models \craig{are} able to encode the \zmss{within-entity context knowledge} from the side information of users and items. \zmrs{Note that \ya{our proposed pre-training scheme} is a general framework to pre-train entity embeddings with side information \ya{in} recommender \ya{systems}. \ya{Once} these embeddings are obtained, they \ya{can} be applied to any existing general representation-based recommenders to enhance their effectiveness and stability.} We deploy our pre-trained \zmss{embeddings} into four existing representative general recommender models, i.e., MF~\cite{rendle2020neural}, NCF~\cite{he2017neural} NGCF~\cite{NGCF19} and LightGCN~\cite{he2020lightgcn} to validate the effectiveness of our \io{proposed} pre-training scheme. 

\thispagestyle{empty}
The contributions of this work can be summarized as follows\footnote{Our source code is available at \url{https://github.com/pretrain/pretrain}.}:
%\inote{TODO: update the GitHub Repo}
\begin{enumerate}[leftmargin=*]
    \item We introduce \craig{a} novel pre-training scheme for leveraging side information \zmrs{by first pre-training the entity embeddings using entity side information and then fine-tuning them using an existing recommender model.}
    \item \zm{We propose two \zmss{pre-training} models using graph neural networks, namely, the \gcn{} model and the \com{} model, which learn entity embeddings based on the single-relational graphs and the multi-relational graphs, respectively, where both types of graphs are constructed from the entity side information.} Both our models can be deployed to fine-tune and enhance existing general representation-based recommender systems.
    \item  \iadh{An extensive empirical} evaluation of our \zmss{pre-training} models -- \craig{through} the fine-tuning \craig{of} four existing recommender models \craig{on} three real-world datasets -- shows that our \zmss{pre-training scheme} can significantly enhance \siwei{those} four models \zmrs{in terms of both the recommendation performance and the model stability.}
    % \item \siwei{An extensive empirical evaluation of our pre-trained models through the fine-tuning of four existing recommender models on three real-world datasets demonstrates that pre-training embeddings using our proposed models can significantly enhance those existing \iadh{models'} performances as well as their stabilities, and that our proposed models can better leverage entity side information compared with state-of-the-art context-aware models using the \textit{integration scheme.}} \inote{duplicate of (3)}
\end{enumerate}

%\todo{Should we add a paragraph outlining the structure of the remaining of the paper? \sws{A outline paragraph is added by Siwei, please check.}}
% \sws{The remainder of this paper is \io{organized} as follows. In \S\ref{sec:related}, we position our proposed model in the literature. \S\ref{sec:notion} and \S\ref{sec:methodology} detail all relevant notions used in this paper, the architecture of our model and \io{describes our} pre-training scheme for recommender models. The experimental setup and the results of our empirical experiments are presented in \S\ref{sec:setup} and \S\ref{sec:result} respectively, followed by \io{some} concluding remarks in \S\ref{sec:concolusion}}.

%\todo[craig]{item 3 is the first mention of fine-tuning}
% Reply by Zaiqiao: we have mention it in the previous paragraph.

% ------------ Related Work ------------
\section{Related Work}\label{sec:related}
In this section, we give a brief introduction about four \io{bodies} of related works: {\em recommendation models, integrating side information for recommendation, graph neural networks} and {\em graph neural recommendation models}.

% \subsection{Recommendation Models}
\noindent\textbf{Recommendation Models.} Recommender systems \iadh{are} essential tools \iadh{that} help resolve the information overload, \iadh{which} have attracted much interest in both academia and industry. Among the various methodologies~\cite{quadrana2018sequence,zhang2019deep} in the evolution of recommender systems, the \em representation-based \em methods, \io{which} learn latent embeddings for users and items have \iadh{been shown} to be the most effective and popular ones in the literature because of their capability of capturing complex user \iadh{preferences} and \iadh{items'} popularity~\cite{he2017neural,rendle2020neural}. Moreover, with the recent development of various deep neural networks, such as Convolution Neural Networks (CNNs)~\cite{yuan2019simple}, Recurrent Neural Networks (RNNs)~\cite{manotumruksa2017deep}, Attention Networks~\cite{kang2018self} and Graph Neural Networks (GNNs)~\cite{NGCF19}, \iadh{more and more} deep neural network-based recommendation models are \craig{being} proposed to cope with various recommendation scenarios, such as temporal-aware~\cite{manotumruksa2017deep}, social-aware~\cite{liu2020heterogeneous}, knowledge-aware~\cite{wang2019kgat} and entity-context-aware~\cite{guo2019exploiting} \iadh{recommendation}, \zm{making the learned embeddings capture more \io{contextual} information}. In particular, \zm{\iadh{in the presence of various critical issues such as the data sparsity issue in the datasets and the cold start problem when  recommending items to new users}}, exploiting the side information of users and items have been widely recognized as an important modelling dimension to address these issues~\cite{guo2019exploiting,chen2018collective}. The main focus of this work is to provide a general scheme to integrate such widely available side information of both users and items into any general representation-based recommender \io{system}.

% \subsection{Integrating Side Information for Recommendation}

\noindent\textbf{Integrating Side Information for Recommendation.} Integrating \iadh{the} side information of entities (such as \craig{each user's age group} and \craig{each item's} textual descriptions) into recommender systems has been widely studied for many years~\cite{chen2018collective,ning2012sparse,xiao2019bayesian}.
Indeed, before the prevalence of deep neural network-based recommendation methods, there \iadh{have been} many variants of the Matrix Factorization-based methods, such as the sparse linear methods with side information (SSLIM)~\cite{ning2012sparse} and the hierarchical Bayesian matrix factorization method~\cite{park2013hierarchical}, which \iadh{adopt the} \emph{integration scheme} that incorporates \iadh{the} entity side information by combining the recommendation loss \iadh{function} with an extra \zms{side information}-aware loss. Even in recent years, this line of research still pervades many works in the literature. For example, the \cm{recently proposed} HIRE~\cite{liu2019recommender} model uses \craig{a} weighted matrix factorization to encode both flat and hierarchical \craig{forms of} side information into \iadh{the} users' and items' representations, \craig{while combining} \io{the} recommendation loss and two \zms{side information}-aware losses. Another line of research \iadh{examined} the integration of side information using deep neural networks, such as \zm{the} stacked denoising auto-encoder~\cite{wang2015relational} and \zm{the} marginalized denoising auto-encoder~\cite{li2015deep}. More recently, many recommendation models have explored using Variational auto-encoders (VAEs)~\cite{chen2018collective,xiao2019bayesian,pang2019novel,wu2020hcvae}, \io{which} jointly encode user ratings and side information when training, \iadh{in order} to overcome the (often) high-dimensionality of side information. \zm{However, most of these methods only consider one type of \craig{relation} from the entity features, namely they treat all the feature columns equally, \iadh{thereby ignoring} the \iadh{variance in the feature types' importance} to the recommendation performance. The HIRE~\cite{liu2019recommender} model considers two types of \iadh{features} (flat and hierarchical), but it needs \iadh{to} explicitly design different objectives \iadh{for each different type of \cm{feature}}.} Moreover, most of these methods adopt the integration scheme, \io{which} needs a trade-off between the recommendation loss \iadh{function} and the \zms{side information}-aware loss, \io{thereby restricting} the model design and \io{making it} hard to deploy into other more effective general recommender systems. \zm{Instead, in this paper, we \iadh{propose} a general framework for pre-training entity embeddings using the entity side information, such that these embeddings can be fine-tuned by an existing representation-based recommender system.}

% \subsection{Graph Neural Networks}
% reorganize this part by: single graph, multi-graph, and heterogeneous graph

\noindent\textbf{Graph Neural Networks (GNNs).} GNNs \craig{are} a powerful framework \cm{for} learning \iadh{representations} \cm{of} graph-\zms{structured} data. \craig{They have} shown superior \iadh{performances} not only in network analysis tasks~\cite{meng2019co,meng2019semi,kipf2017semi,velivckovic2017graph} but also in \io{other} \io{domains}, such as \io{natural} language processing~\cite{zhu2019graph}, recommender systems~\cite{NGCF19} and molecular design~\cite{you2018graph}. Most of the existing GNNs focus on learning representations of nodes on \em single-relational graphs \em that contain nodes and relations of a single type. The Graph Convolutional Network (GCN)~\cite{kipf2017semi} \iadh{model} and its variants~\cite{hamilton2017inductive,velivckovic2017graph,meng2019semi} \craig{are} probably the most popular models for handling this type of graphs. As a matter of fact, real-world \craig{graphs} usually come with \io{multiple} types of relations, i.e.\ relations \io{associated} with labels such as friendships and co-worker relationships, \craig{thereby forming} a modeling data structure widely known as \em multi-relational graphs\em~\cite{nickel2011three,vashishth2019composition,schlichtkrull2018modeling}. Because \iadh{a} multi-relational graph contains more comprehensive information and rich semantics, it has been widely used in many \craig{tasks that mine} knowledge graphs, such as lexical word networks~\cite{sun2018rotate,zhang2019quaternion} and biomedical knowledge  graphs~\cite{chang2020benchmark}. Relational-GCN~\cite{schlichtkrull2018modeling} and \craig{Compositional}-GCN~\cite{vashishth2019composition} are two generalizations of the GCN model~\cite{kipf2017semi} for handling multi-relational graphs. In this paper, we propose to use graph neural networks to capture the within-entity context knowledge from the side information of users and items, where both the single-relational graphs and the multi-relational graphs are \zms{explored for pre-training \io{the} entity embeddings}.

\thispagestyle{empty}
% \subsection{Graph Neural Recommendation Models}

\noindent\textbf{Graph Neural Recommendation Models.} Inspired by the wide variety of applications in many fields, graph neural networks have \iadh{also} been widely applied in many recommendation models, showing promising results \io{on} \craig{various} recommendation benchmarking datasets~\cite{NGCF19,he2020lightgcn,liu20202,hu2020graph,wang2019kgat,yu2020enhance,liu2020basconv,chen2020revisiting, liu2020heterogeneous,wu2020graph}.
Most of these methods~\cite{NGCF19,chen2020revisiting,he2020lightgcn,zhang2020stacked} consider the user-item interactions in \craig{a} recommendation task as a user-item graph, and \iadh{use} the power of GNNs to capture the higher-order \cm{dependencies} among the entities, \zm{resulting in \iadh{enhanced recommendation \io{performances}}.} % thereafter the recommendation performance can be improved.
Many \io{prior} works also exploited using GNNs to model various \zmss{cross-entity contextual knowledge (i.e.\ interactions across different types of entities)}~\cite{hu2020graph,liu20202,wu2020graph, wei2019mmgcn} and \zmss{within-entity contextual knowledge (i.e.\ relations within a type of entity)}~\cite{liu2020heterogeneous,liu2020basconv,zheng2020price,yu2020enhance}.
For example, both the A$^2$-GCN~\cite{liu20202} \craig{and} the GCM-C models~\cite{wu2020graph} deploy the GNNs to learn the representations \iadh{for} the graph constructed by the context edges between the user-item pairs. \craig{However,} they are unable to learn the \zms{within-entity contextual knowledge}. \zms{Another approach proposed by \citet{hu2020graph} deploys GNNs for the news recommendation task to explicitly model the interactions among users, news and latent topics by constructing a user-news-topic heterogeneous graph, which can also \io{be} seen as a cross-entity \io{context} modeling approach.} Some special within-entity contexts, such as social information~\cite{fan2019graph,yu2020enhance}, text semantics~\cite{liu2020heterogeneous} and basket context~\cite{liu2020basconv} have also been modeled by GNNs. %\io{However,} they are unable to \iadh{generalize} into other general recommendation models.
\zms{In particular, \io{as far as we know}, the Multi-GCCF model~\cite{sun2019multi}, which also deploys the multi-relational graph neural model into recommendation systems, is the most relevant work to our proposed models \cm{- a} more detailed discussion comparing our \cm{proposed} model with this model is provided in \S\ref{sec:discus}.} % \inote{should we not then say how we differ from Multi-GCCF or is this answered in the next paragraph?} Zaiqiao: Yes we discuss the different in section 4.5
%\cm{as far as} we know

\zm{However, all \craig{of} the existing recommendation models discussed above \craig{are} either unable to effectively leverage entity side information, or \craig{are} unable to effectively encode \iadh{the} \zmss{within-entity contextual knowledge} from the side information in a general \iadh{manner}. Therefore, in this paper, we provide a general framework for \iadh{leveraging} the entity side information using a novel graph neural pre-training scheme, which is detailed in the following sections.}
% pre-train models
% provide a figure to illustrate the different pretrain schemes

% ------------ Our Model ------------

% \section{PRELIMINARIES}
% In this section, we first introduce notations we used and the recommendation task to be addressed in this paper. Then, we briefly describe two graph neural networks that are used in this work.

\thispagestyle{empty}
\section{Notations and Task Definition}\label{sec:notion}
% Throughout this paper, regular letters are used to denote \iadh{scalars}, while calligraphy typeface alphabets are used to denote sets.  Matrices and vectors are denoted by bold letters with uppercase letters representing
% matrices and lowercase letters (or uppercase letters with a subscript) representing vectors.

%The general formulation of the recommendation task \iadh{consists in learning} the latent representation vectors (i.e. embeddings) of \iadh{both} the user \iadh{and} item entities, for a given interaction/rating matrix, over the two types of entities, such that the learned representation vectors can \iadh{be} used to predict the preference scores for \iadh{the} user-item pairs. %We let $\set{U} = \{u_1,u_2,\cdots ,u_n\}$ to be the set of $n$ users and $\set{V} = \{v_1,v_2,\cdots ,v_m\}$ be the set of $m$ items. 
\zms{Let $\mat{R}\in \mathbb{R}^{m\times n}$ be the user-item feedback matrix of $n$ users and $m$ items. \zms{\io{In general,} the entries of $\mat{R}$ can be \io{either binary-valued} (i.e. \io{through} implicit feedback such as the click and purchase behaviours) or real-valued (i.e. \io{through} explicit feedback such as the \io{rating} scores). \io{However,} not all \io{of} the values can be observed and the observed values may also contain noise. While the observed entries at least reflect \io{the real} users' \io{interests} on items, most of the entries are \io{typically} missing due to the natural \io{sparsity} of negative feedback.} A typical recommendation task \io{aims}, given $\mat{R}$, to predict the preference scores for all the user-item pairs \zms{(both the missing and observed pairs)}. Note that \io{the} predicted preference scores can be used to perform both the rating prediction/matrix completion and item ranking tasks. To address such recommendation \cm{tasks}, the general representation-based methods commonly learn latent embeddings for users and items such that these latent entity embeddings can be used to calculate the preference scores. \io{In particular,} \io{such methods} try to learn the following mapping function:}
\begin{equation}
    \mat{R} \xrightarrow{} \mat{U}, \mat{V},
\end{equation}
where $\mat{U} \in \mathbb{R}^{n\times d}$ and $\mat{V} \in \mathbb{R}^{m\times d}$ are the learned latent embeddings with $d$ being the embedding dimension.

In real-world recommender systems, users and items are often associated with features, also called side information, such as the \iadh{age groups} of users and the textual \iadh{descriptions/names} of items. \zms{Such side information \iadh{about} entities can be used to learn the \zm{within-entity context knowledge} \iadh{to further enhance} the recommendation performance.}
%\inote{split sentence here, are you talking about side information being an important modelling dimension or the recommendation performance? sentence suggests you mean the latter which is incorrect}. 
We let $\mat{F}_{u}\in\mathbb{R}^{n\times k_1}$ and $\mat{F}_{v}\in\mathbb{R}^{m\times k_2}$ denote the feature vectors of users and items, with $k_1$ and $k_2$ \cm{being their respective number of features}. %being their feature numbers respectively. 
Then, the \io{corresponding} recommendation task \io{consists in learning} a mapping function \iadh{as follows}:
\begin{equation}
    \mat{F}_{u}, \mat{F}_{v}, \mat{R} \xrightarrow{} \mat{U}, \mat{V}.
\end{equation}
In order to facilitate the description of graph neural networks, we use $\set{G}=(\set{V},\set{E},\mat{A}, \mat{X})$ to denote a \emph{single-relational graph} (simply called \iadh{a} graph) with $n$ nodes, where $\set{V}$ and $\set{E}$ are the sets of nodes and edges, respectively. $\mat{X}\in\mathbb{R}^{n\times k}$ is the feature matrix of nodes, \zms{where $k$ is the number of node features.} $\mat{A}\in \mathbb{R}^{n\times n}$ is the adjacency matrix often used to represent the matrix form of edge weights. We also denote a \emph{multi-relational graph} by $\set{G}^{'}=(\set{V},\set{E}^{'},\set{R}, \mat{X})$, where each edge $(u,v,r) \in \set{E}^{'}$ \cm{has an associated} relation type $r \in \set{R}$\footnote{In this paper, we only consider the unweighted case of a multi-relational graph, hence no adjacency matrix is needed to represent the edge weights.}.

% In this paper, we assume all the associated side information are \textit{categorized} features, i.e. each entity contains sparse bag-of-words feature vectors.  Although in real-world there are many real-valued feature associated with users and items, we can easily categorized them by grouping values into buckets. We explore two types of graph, i.e. \emph{single-relational graph} and \emph{multi-relational graph},  extracted from entity features to learn graph neural network-based representation.
%  \emph{single-relational graph}  is a type of graph that regards all the edges in the graph as a \emph{homogenous} type. We denote such single-relational graph as $G=(\set{V}, \set{E})$ with nodes (entities) $v_{i} \in \set{V}$ and edges $\left(v_{i}, v_{j}, w_{ij}\right) \in \set{E}$, where $w_{ij}$ is the edge weight.  \emph{multi-relational graph} is a type of graph that all the edges in the graph are associated with a edge type. We denote such multi-graph as $G=(\set{V}, \set{E}, \set{R})$ with nodes (entities) $v_{i} \in \set{V}$ and labelled edges $\left(v_{i},r, w_{ij}, v_{j}\right) \in \set{E}$, where $r\in \set{R}$ is a edge type.  We can easily extract both the \emph{homogeneous} and \emph{heterogeneous} links between them and construct graphs for each type entities, which \craig{are} detailed in Section \ref{sec:pre_train}.

\section{METHODOLOGY}\label{sec:methodology}
In this section, we \zmrs{first introduce our graph neural pre-training scheme for recommender systems} \io{(\S\ref{ss:framework})}, and detail \iadh{the} pre-training \zmrs{schemes on single-relational feature graphs \io{(\S\ref{sec:pre_train}) and multi-relational feature graphs (\S\ref{ss:pre-train-multiple})}, respectively. \ya{We then} describe the fine-tuning process using the \io{existing} \io{recommender models} \io{(\S\ref{ss:fine_tune}) \ya{before discussing the complexity of the proposed scheme and its contribution in \S\ref{ss:complexity} and \S\ref{sec:discus}, respectively}}}. 
%\inote{a little unclear - "as well as the fine-tuning process with existing .. models"
%  -- do you mean when applied to existing recommender models?}
 
\subsection{The Pre-training Framework for Recommendation} \label{ss:framework}
\begin{figure*}
    \centering
   \includegraphics[width = 140mm]{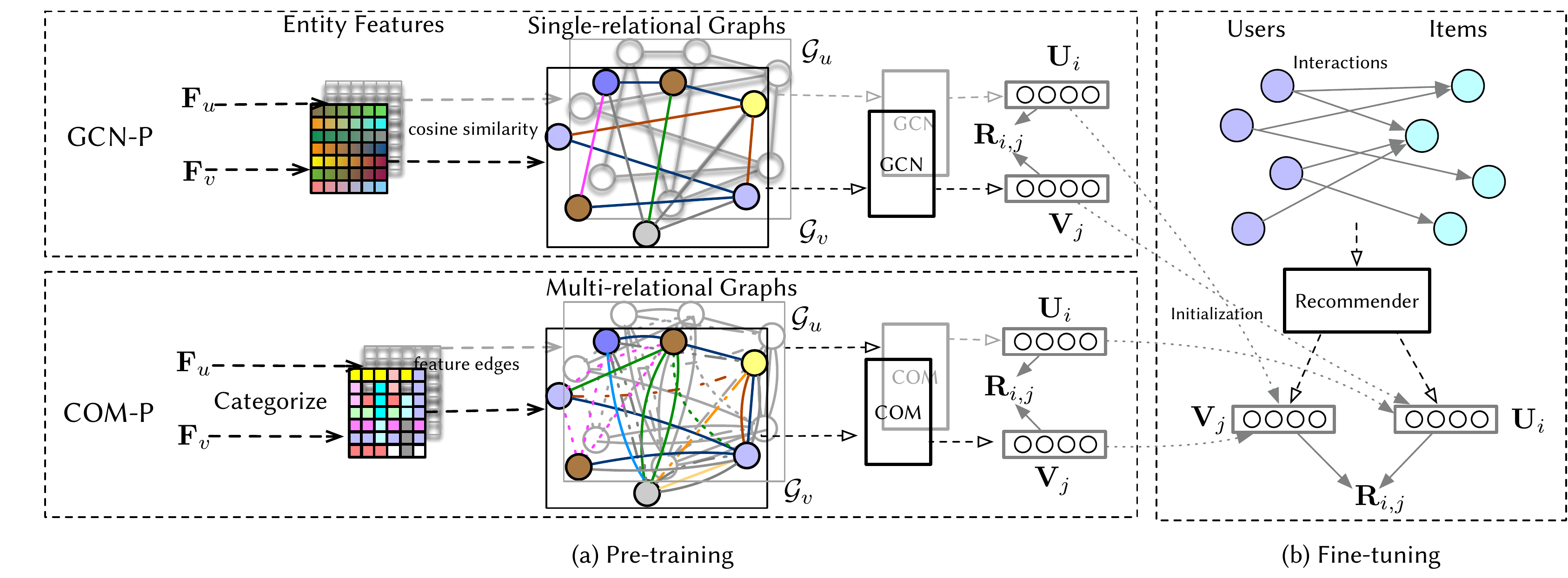}
   \caption{\label{fig:overview}\zmrs{\cmrs{An overview of our} graph neural pre-training scheme}. The \gcn{} pre-training model constructs two single-relational graphs based on the feature cosine similarities of \iadh{the} entity pairs, and pre-trains the embeddings of entities by using the Graph Convolutional Network (GCN)~\cite{kipf2017semi} model.  The \com{} pre-training model constructs two multi-relational graphs based on the categorized feature edges of \iadh{the} entities, and pre-trains the embeddings of entities by using the Composition-GCN (COM)~\cite{vashishth2019composition} model.}
\end{figure*}

\thispagestyle{empty}
As \iadh{mentioned} in the introduction section, \iadh{an} appropriate initialization of \io{the} \zms{entity} embeddings is critical to help recommenders \io{learn} a good local minimal solution for \zms{the existing recommender systems}. In this paper, we propose to learn \zms{such an initialization of entity embeddings} by exploiting knowledge from the entity side information.
To this end, we propose a general pre-training framework for leveraging the entity side information using graph neural networks.
The overall framework is illustrated in Figure \ref{fig:overview}. Our pre-training framework consists of two processes: \cm{\emph{pre-training} and \emph{fine-tuning}. During \emph{pre-training}}, a graph neural network is used to learn an initialization of the entity embeddings based on \zms{both the entity side information and \io{the} \zms{feedback} matrix $\mat{R}$}. \io{On the other hand,} in the \emph{fine-tuning} process, an existing recommendation model \iadh{leverages} the pre-trained embeddings as \io{an} embedding initialization and \iadh{fine-tunes} these embeddings by using the \zms{feedback} matrix $\mat{R}$ only.

%\zms{GNNs, as a type of powerful neural model for learning entity} \iadh{representations} by smoothing features over the entities graph~\cite{kipf2017semi,vashishth2019composition,schlichtkrull2018modeling}, have been widely applied into the recommendation domain recently, such as NGCF~\cite{NGCF19} and LightGCN~\cite{he2020lightgcn}, achieving state-of-the-art performances for recommendation. 
% \inote{Todo: Unify two concepts for the whole paper: within-entity context and cross-entity context}
\zms{In order to learn \io{the} user-item preferences from the \zms{cross-entity contextual interactions} between users and items, many recent models, such as NGCF~\cite{NGCF19} and LightGCN~\cite{he2020lightgcn}, have explored \io{encoding} the collaborative signal from the graph-structure interactions, showing promising \io{performances}.
However, these methods only investigate \iadh{the} mutual interactions between users and items, ignoring the \zmss{within-entity contextual knowledge}, \zms{i.e. the collaborative signal within each type of entities, which can be acquired from the entity side information.}  To capture such \zmss{within-entity contextual knowledge},} we propose to \io{use} GNNs to pre-train \iadh{the} entity representations using the side information of both the users and items, so that the independent knowledge of each \io{type} of entities can be captured from \iadh{the} entity relations. Hence, extracting the relations from \iadh{the} entity side information is \zmrs{a}
% \inote{claim. it could be softened by saying it is "a crucial step"}
crucial step for the pre-training process, since it determines how much information we can obtain from \iadh{the} entity features and how important such information can help to improve a recommendation model. To extract user-user and item-item relationships \zms{from their entity features (i.e. \io{the} users' and items' respective features), we propose to build two different types of feature graphs}, i.e. the single-relational graphs and the multi-relational graphs, by constructing both the \emph{homogeneous} and \emph{heterogeneous} links between \iadh{the} entity pairs of each entity type, respectively. \iadh{We then} explore how entity relations from various entity features affect the recommendation performance.

\subsection{Pre-training on Single-relational Graphs}
\label{sec:pre_train}
\thispagestyle{empty}
We construct two single-relational graphs, i.e. $\set{G}_u$  and $\set{G}_v$,  from \iadh{the} features of users and item respectively,  by \iadh{considering} the similarities between users and items as the homogeneous edges, and calculate the edge weights \io{using} the cosine similarities between each pair of entities. For example, we construct a \emph{user single-relational graph} $\set{G}_u = (\set{V}_u,\set{E}_u, \mat{A}_u, \mat{X}_u)$ by \zms{taking all the users as \io{the} set of nodes $\set{V}_u$ and all the user pairs as \io{the} set of edges $\set{E}_u$ in the graph. For each $(i,j) \in \set{E}_u$, we calculate the edge weight $\mat{A}_{ij}$ using the cosine similarity of their feature vectors (i.e. $\mat{F}_{{u}_i}$ and $\mat{F}_{{u}_j}$):}
$\mat{A}_{ij}  = \frac{\mat{F}_{{u}_i} \cdot \mat{F}_{{u}_j}}{\|\mat{F}_{{u}_i}\|\|\mat{F}_{{u}_j}\|}$.
\zms{$\mat{X}_u\in \mathbb{R}^{n\times d}$ is \io{an} initial node feature matrix of the graph, the values of which are initialized from the uniform distribution $\mathcal{U}(-0.01, 0.01)$}. For brevity, in the following,  we only describe the encoding process for \iadh{the} user single-relational graph $\set{G}_u$, since the \emph{item single-relational graph} $\set{G}_v$ is constructed and processed in a similar \io{fashion}.

\thispagestyle{empty}
\zmss{\textbf{The \gcn{} model.}} To obtain the \zms{pre-trained} embeddings of entities (i.e. users and items) and \io{to} exploit the potential correlation among entities based on their single-relational graphs, \swss{three} GCN layers~\cite{kipf2017semi} are applied to encode the entity embeddings according \io{to} their relations.
%GCN~\cite{kipf2017semi} is one of the most classical graph neural networks, \iadh{which} generalizes the definition of convolution from the regular grid data to the graph-based data.
The key point of GCN is to propagate \iadh{the} feature information through neighbourhoods of nodes \io{in} each iteration during training.
Specifically, given a graph $\set{G}_u$, the GCN model adopts the following propagation rule:
\begin{equation}
\label{eq:gcn}
\mat{H}_u^{(l)}=f\left(\hat{\mat{A}_u} \mat{H}_u^{(l-1)} \mat{W}_u^{(l-1)}\right),
\end{equation}
where $\hat{\mat{A}}_u=\widetilde{\mat{D}}^{-\frac{1}{2}}(\boldsymbol{A}_u+\mat{I}) \widetilde{\mat{D}}^{-\frac{1}{2}}$ is the symmetric normalized adjacency matrix\footnote{$\widetilde{\mat{D}}$ is defined as $\widetilde{\mat{D}}_{i i}=\sum_{j}(\mat{A}_u+\mat{I})_{i j}$, where $\mat{I}$ is the identity matrix.}, $\mat{W}_u^{(l)}$ is the weight matrix of the $l^{th}$ layer, and \zmss{${f}$ denotes \io{an} activation function (e.g. the ReLU function)}. $\mat{H}_u^{(l)}$ is the hidden node representation in the $l_{th}$ layer with $\mat{H}_u^{(0)} = \mat{X}_u$. \zm{To facilitate the \iadh{later} description of our \iadh{proposed} model,} \zmss{\eq{}\eqref{eq:gcn}} can \iadh{also} be formulated \io{in} \zmss{the} message passing form~\cite{xu2018powerful,vashishth2019composition}:
\begin{equation}
\label{eq:layer}
\mat{H}_{u_i}^{(l)}=f\left(\sum_{(u_i,u_j) \in \set{E}_u\cup(u_i,u_i)}\mat{W}_{u}^{(l-1)}\mat{H}^{(l-1)}_{u_j}\right),
\end{equation}
where $\mat{W}_{u}^{(l-1)}$ is the layer-wise parameter and here we only consider \io{only} the undirected relation and \io{the} self-loop relation. The final output embeddings of the maximum depth in \iadh{the} GCN layers, i.e. $\mat{U} =\mat{H}_{u}^{(l)}$, are the pre-training embeddings to be \iadh{fed} into \iadh{the} pre-training loss function.  In this paper, we set the maximum layer depth \iadh{to} $3$, \iadh{since a} higher depth \io{might} cause \io{an} over-smoothing issue~\cite{zhou2020towards,oono2020optimization}. \zmrs{Similarly}, the item embeddings are obtained by $\mat{V} = \mat{H}_{v}^{(l)}$, where $\boldsymbol{H}_{v_{i}}^{(l)}=f\left(\sum_{\left(v_{i}, v_{j}\right) \in \mathcal{E}_{v} \cup\left(v_{i}, v_{i}\right)} \boldsymbol{W}_{v}^{(l-1)} \boldsymbol{H}_{v_{j}}^{(l-1)}\right)$, which is aligned with \eq{} \eqref{eq:layer}.

 \textbf{Pre-training loss.} For most of the graph neural networks, the embedding can be learned by the reconstruction of the graph structure or the labels of the corresponding entities~\cite{kipf2017semi}. In our pre-training scheme, to make the representations of \iadh{the} entities that capture the potential correlation between entities also \craig{usable} for recommendation, we train the model with a rating/interaction-based loss. Specifically, with the embeddings of both users and items from the GCN layers, we construct our pre-training loss \zms{using a} Binary Cross-Entropy (BCE) loss function~\cite{he2017neural}:
% illustrate the formulations of the two losses, if we have enough space
%\begin{equation}
%	L_{BPR}=-\sum_{{(u,i)}
%		\in \mathbf{R}}  \sum_{{(u,i)}
%		\notin \mathbf{R}}  \ln \sigma\left(\hat{y}_{u i}-\hat{y}_{u j}\right)+\lambda\left\|\mat{\Theta}\right\|^{2},
%\end{equation}
\begin{equation}
	\label{eq:loss}
	\set{L}_{BCE}=- \sum_{{\mat{R}_{i,j}}
		\in \mat{R}} \mat{R}_{i,j}\cdot \log \left(\hat{\mat{R}}_{i,j}\right)+\left(1-\mat{R}_{i,j}\right) \cdot \log \left(1-\hat{\mat{R}}_{i,j}\right)+\lambda\left\|\mat{\Theta}\right\|^{2},
\end{equation}
where $\hat{\mat{R}}_{i,j}$ is the predicted score calculated by the embedding dot product $\hat{\mat{R}}_{i,j}=\mat{U}_i^{\top}\mat{V}_j$, and $\mat{\Theta}$ is a regularization term parameter of embeddings. Two explicit entity biases are also used for calculating the scores, \zms{following}~\cite{rendle2020neural}. We \io{refer} \iadh{to} our pre-training model using \io{the} GCN layers on single-relational graphs \zmrs{as} \gcn{}.

\subsection{Pre-training on Multi-relational Graphs} \label{ss:pre-train-multiple}

In the single-relational graphs, different types of features associated with users and items contribute equally to the latent correlations between entities. However, in real-world recommender data, \iadh{the} features of users and items are typically heterogeneous, with different types of features \zm{having} different \cm{usefulness for enhancing} a recommender's performance. For example, users are also associated \io{with} different types of features (e.g. age, gender and education level), which clearly characterize different aspects of \io{the} users' \iadh{preferences}~\cite{liu2019recommender}. It is hard to distinguish \io{the} different \io{feature types' importance} from the entity correlations constructed by \iadh{the} cosine similarity, which treats different types of features with equal weights. Hence, \zms{instead of propagating the feature messages between entities} \zm{through} the homogeneous edges,
%\inote{not sure about this sentence. what is it we are assuming?}
\zms{in this \craig{section} we \iadh{aim} to learn to propagate feature \cm{messages} through different feature types of \zmss{the entities} by using the multi-relational graphs.}

Since the entity \iadh{features} can be real-valued, we need to categorize \iadh{such} real-valued features into some groups, such that all the features of entities are \zms{sparsely} categorized. \io{Next}, we can regard each feature category value as an edge type, and create an edge of this type between a pair of entities if they share the same feature value. In particular, to extract the heterogeneous relations between entities, we construct two multi-relational graphs for \iadh{the} users and items,  i.e. $\set{G}^{'}_u$  and $\set{G}^{'}_v$, respectively. \zms{For example, to construct the \emph{user multi-relational graph} $\set{G}^{'}_u=(\set{V}_u,\set{E}^{'}_u, \set{R}_u, \mat{X}_u)$, we take all the users as \io{the} set of nodes $\set{V}_u$, and \io{the} feature category values as the set of relations $\set{R}_u$ in the graph. For any pairs of nodes $i,j \in \set{V}_u$, we create an edge if the two users share \swss{the} same feature category value (e.g. \iadh{if both are} in \io{the} \zms{age} group 25-35).} The \emph{item multi-relational graph} $\set{G}^{'}_v$ is constructed in a similar \io{fashion}.

To handle different edge types in both constructed multi-relational graphs, many multi-relational graph embedding models are available for our pre-training task, among which Relational-GCN~\cite{schlichtkrull2018modeling} and Composition-GCN~\cite{vashishth2019composition} are the most effective ones from the literature. \cm{In this paper, we apply the Composition-GCN model, since our own initial experiments found \zmss{that it is} more efficient and more effective in \craig{a} multi-relational graph representation learning task than the Relational-GCN model.}
%~\cite{vashishth2019composition} 

%according to our \io{own} empirical \iadh{results}, the \craig{Compositional}-GCN model~\cite{vashishth2019composition} \craig{has been shown to be} both more efficient and more effective in \craig{a} multi-relational graph representation learning task than the Relational-GCN model~\cite{schlichtkrull2018modeling}.

%\textbf{Composition-GCN.} 
%The composition-based multi-relational graph convolutional network~\cite{vashishth2019composition} is a recently proposed graph neural network for handling multi-relational graphs. 

\thispagestyle{empty}
\zmss{\textbf{The \com{} model.}} Given a multi-relational graph, e.g. \zmss{the user multi-relational graph} $\set{G}^{'}_u=(\set{V}_u,\set{E}^{'}_u, \set{R}_u, \mat{X}_u)$, the Composition-GCN first extends $\set{E}^{'}_u$ and $\set{R}_u$ with the corresponding inverse edges and relations:
\begin{align}
	\hat{\set{E}}_u^{\prime}=&\set{E}_u \cup\left\{\left(v, u, r^{-1}\right) \mid(u, v, r) \in \set{E}_u\right\} \cup\{(u, u, \top) \mid u \in \set{V}_u),\nonumber\\
	\hat{\set{R}}_u=&\set{R}_u \cup \set{R}^{i n v}_u \cup\{\top\},
\end{align}
where $\set{R}^{i n v}_u=\left\{r^{-1} \mid r \in \set{R}_u\right\}$ denotes the inverse relations \zmss{(i.e. $(v, u, r^{-1}) = (u, v, r)$)} and $\top$ indicates the \io{self-loop}. Then, \iadh{the} node embeddings are propagated through edges based on the following \iadh{propagation} rule:
\begin{equation}
\label{eq:com}
	\mat{H}^{(l)}_{u_i}=f\left(\sum_{(u_i, u_j, r) \in \hat{\set{E}}_u^{\prime}} \mat{W}^{(l-1)}_{\lambda(r)} \phi\left(\mat{H}^{(l-1)}_{u_j}, \mat{Z}^{(l-1)}_{r}\right)\right),
\end{equation}
where $\boldsymbol{W}^{(l-1)}_{\lambda(r)} \in \mathbb{R}^{d_{1} \times d_{0}}$ is a relation-type specific parameter, $\mat{Z}_{r}=\sum_{k=1}^{b} \alpha_{k r} \mat{B}_{k}$ is the relation embedding \zms{with $\left\{\mat{B}_{1}, \mat{B}_{2}, \ldots, \mat{B}_{b}\right\}$ being a set of learnable basis vectors and $\alpha_{k r}$ \swss{is} \zms{the basis-specific} learnable scalar weight.  $b$ is a hyperparameter corresponding to the number of basis vectors. In \eq{}\eqref{eq:com}, $\phi$ is a  composition operator defined by: $\phi\left(\boldsymbol{e}_{s}, \boldsymbol{e}_{r}\right)=\boldsymbol{e}_{s}-\boldsymbol{e}_{r}$, which is inspired \cm{by} the TransE model~\cite{bordes2013translating}.}

Then the output embeddings of the final layer are taken as pre-training embeddings (i.e. $\mat{U} =\mat{H}_{u}^{(l)}$ ). The item embeddings are obtained similarly to \eq{}\eqref{eq:com}, and the BCE loss (\eq{}\ \eqref{eq:loss}) is also applied to define the pre-training loss. We refer \iadh{to} our pre-training model using the Composition-GCN layers on multi-relational graphs \ya{as} \com{}.

\subsection{Fine-tuning with Existing Recommender \iadh{Models}} \label{ss:fine_tune}
Most of the modern recommenders are trained based on gradient-based optimization methods, which can only \io{obtain} locally-optimal solutions. Due to the non-convexity of their objective function\swss{s}, parameter initialization plays an important role for the convergence and \iadh{performances} of these recommendation models~\cite{he2017neural,ebesu2018collaborative}. Most of the existing recommendation models initialize their embeddings from a uniform distribution~\cite{wan2018representing}, a normal distribution~\cite{rendle2020neural} or the \zmss{Xavier uniform distribution}~\cite{he2020lightgcn,NGCF19}. Due to the randomness of generated embeddings and the lack of prior knowledge, these models often fall into some poor locally-optimal solutions, resulting \zms{in} high instabilities, as illustrated in \S\ref{sec:Introduction}.
%the \io{Introduction} section. 

To address this issue, we propose to initialize the entity embeddings of an existing recommendation model from the output embeddings of our \zmss{pre-training} models, then \io{we} further fine-tune these embeddings with the recommendation model's own parameter optimizer. Specifically, we first pre-train both the embeddings of users and items by one of our proposed models (i.e. \gcn{} or \com{}) until convergence, then feed these pre-trained embeddings into an existing recommendation model as the parameter initialization \zms{to train the recommendation model} with the interactions/ratings only. To investigate the effect of different types of entity relations \iadh{on} the recommendation performance, we experiment with both \io{of} our two proposed pre-training  models, and fine-tune the pre-trained embedding by four existing representative recommendation models, i.e. the Matrix Factorization (MF)~\cite{rendle2020neural}, Neural Collaborative Filtering (NCF), Neural Graph Collaborative Filtering (NGCF)~\cite{NGCF19} and LightGCN~\cite{he2020lightgcn} models, \iadh{as will be detailed in the remaining sections}.

\thispagestyle{empty}
%which will be detailed in the later section.
\subsection{Complexity Analysis} \label{ss:complexity}
\zmrs{The complexity of our pre-training scheme highly depends on the complexity of the two graph neural models, i.e. the GCN model holding a complexity of $O\left(ld|\mathcal{E}|+l|d^{2}|\mathcal{V}\right)$~\cite{kipf2017semi} and the Composition-GCN model holding a complexity of $O\left((ld^{2}+bd+b|\mathcal{R}|)|\mathcal{E}|\right)$~\cite{vashishth2019composition}, where $l$ denotes the number of layers, \ya{$b$ is the number} of learnable basis vectors, $d$ is the embedding dimension, and $|\mathcal{R}|$ is the number of relation types. The number of edges $|\mathcal{E}|$ in the entity graphs (multi-relational graphs) \ya{typically} accounts for the largest complexity in both models, and if the entities contain dense features, the number of edges $|\mathcal{E}|$ could be much larger than the number of interactions (e.g. the generated user multi-relational feature graph of the Movielens-1M dataset contains \swrs{around 2.7 million edges}). Some variants of these neural models, such as FastGCN~\cite{chen2018fastgcn} and ClusterGCN~\cite{chiang2019cluster}, \ya{might} enhance the efficiency of our scheme \ya{but at the possible} cost of reducing effectiveness. However, once the entity embeddings are pre-\cmrs{trained}, they can be reused and fine-tuned by many existing recommenders to enhance their effectiveness and stability. \ya{We leave to future work, the investigation of the impact and added-value of these existing variants}.}

%#todo: provide the psudo-codes for the whole process?

\subsection{Discussion}
\label{sec:discus}
\thispagestyle{empty}
The idea of pre-training embeddings for recommender systems \iadh{has} already been investigated in the \iadh{literature}. For example, to avoid saddle points and \zmrs{poorly performing} local minima, both NCF~\cite{he2017neural} and CMN~\cite{ebesu2018collaborative} apply the Generalized Matrix Factorization (GMF) as a pre-training model to initialize \iadh{the} embedding weights of users and items. In particular, the embedding vectors of users and items in GMF are simply obtained by training from the weighted output of the embedding dot product using the interaction matrix:
\begin{equation}
\hat{\mat{R}}_{u,v}=\mat{w}^{\top} \phi\left(\mat{U}_{u} \odot \mat{V}_{V}\right),
\end{equation}
 where $\odot$ denotes the element-wise product of vectors, $\phi$ is \iadh{an} activation function and $\mat{w}^{\top}$ is the trainable parameter. However, these models are unable to leverage the \zms{within-entity context knowledge} from the entity features. We note that both our pre-training models using the graph neural networks \iadh{can be seen as} generalizations of the GMF model, since if we remove all the links in both the constructed single-relational graphs and multi-relational graphs, and set the maximum depth of layers to be 1, \io{then} both \io{of our} pre-training models will \zms{be equivalent to} \io{a} GMF model. \zmrs{More recently, \citet{hao2021pre} \ya{exploited} how to use GNN to conduct pre-training for downstream tasks, inspired by the GNN pre-training~\cite{hu2020gpt}. However, their proposed models only leverage the graph structure, which lacks \ya{the} ability to incorporate heterogeneous auxiliary information \ya{about the} entities, \ya{unlike} our \ya{proposed} scheme.} 
 %\inote{I think the transition here could be more smooth- can we think of another way to link these 2 sentences?} In addition,}
 
\iadh{In relation to} the graph neural-based recommender \iadh{systems}, the most relevant works to our model (\com{}) is the Multi-GCCF model~\cite{sun2019multi}, \zm{which, similarly, considers the user-to-user and item-to-item relations as graphs and uses a graph convolution network to train the embeddings of the two entities.  However, the entity graphs in Multi-GCCF are constructed from the rating/click matrix, rather than \iadh{from} the entity side information. Hence, both the homogeneous and heterogeneous relations from \iadh{the} entity side information \craig{cannot be} captured by the Multi-GCCF model.}

 %Todo: Discuss Which types of representation-based recommendation model cannot be applied in the framework.
\thispagestyle{empty}
\section{Experimental Setup}\label{sec:setup}
% In this section, we \iadh{first introduce the research questions we aim to answer in this paper. Next, we present} the datasets \iadh{used} for conducting \iadh{the} experiments \iadh{as well as} the relevant pre-processing procedures to prepare the \iadh{datasets} including all interaction data and different types of side information. \iadh{Finally}, we present the \iadh{experimental} settings and \iadh{describe the used} baselines.

\subsection{Research Questions}
\label{sec:rqs}
We aim to answer  the following research questions:
\begin{enumerate}[label=\textbf{RQ\arabic*},leftmargin=*]
    \item \iadh{Do} our \iadh{pre-trained} models help existing recommendation systems \iadh{obtain} better \iadh{performances}?
    \item \iadh{Do} our \iadh{pre-trained} models outperform the \io{existing} state-of-the-art recommenders \cm{that use} side information?
    \item \zms{Are the performance improvements gained through our \zmss{pre-training} models due to the within-entity knowledge?}
    \item \iadh{Does} the pre-training process help to improve the stability of the existing models?
    % \item How \iadh{do} the hyperparameters\inote{should this RQ be more specific, ie clear that the hyper parameters are THOSE OF the pre-training models?} affect the \iadh{performances} of our pre-training models? 
    \item \zmrs{How do the \ya{embeddings} dimension and different cut-offs of the ranking \ya{affect} the recommendation performances of our pre-trained models?}
\end{enumerate}

\thispagestyle{empty}
\subsection{Datasets}
To evaluate the effectiveness of our proposed pre-training models, we use three datasets, \io{namely} Foursquare\footnote{\url{https://sites.google.com/site/yangdingqi/home/foursquare-dataset}}, Movielens-1M\footnote{\url{https://grouplens.org/datasets/movielens/}}, and \sws{Epinions}\footnote{\url{http://cseweb.ucsd.edu/~jmcauley/datasets.html}}. The statistics of all three datasets are given in Table~\ref{tab:data_stats}. For the Movielens-1M dataset, the \iadh{users' features \craig{(i.e.\ side information)} are}  ``gender'', ``age'' and ``occupation", while the \iadh{items' features are} the 18 different genres. For the Foursquare dataset, we use \iadh{the} tags provided by online users as features \iadh{for} the restaurants, while we represent each user as a bag-of-words feature vector from his/her own reviews (stopwords are removed). \sws{Similarly, \io{for the Epinions dataset,} we represent both users and items with bag-of-words feature vectors from \io{their associated} reviews \swss{with stopwords removed, selecting the 10 most frequent words to represent both users and items.}.}
%\inote{with stopwords removed?}.
\siwei{Among the three datasets, there is only one real-valued feature, \cm{namely} ages in the Movielens-1M dataset, which needs to be pre-processed into one-hot representations by a categorization operation.}
% All the features of the three datasets are preprocessed into \craig{one-hot representations} by a categorization operation. 
Specifically, the users' age feature in the Movielens-1M dataset \iadh{is categorized} into 8 age groups, each with a step of 10 years. Then \iadh{an} 8-dimensional vector \iadh{is} used to represent \iadh{the} users' age feature.
%\inote{Check Epinions does not have real-valued features Siwei:Epinions only reviews, no real-valued features}
The single-relational graphs per dataset are constructed by the cosine similarity of the features between each entity pairs, \iadh{while} the multi-relational graphs are constructed by the category values.
 
 % \inote{unclear if the last sentence applies to all features or just age}
 % Zaiqiao: For all the features.

\begin{table}
\centering
\caption{\label{tab:stat_all}Statistics of the datasets.}
\resizebox{0.99\columnwidth}{!}{
\begin{tabular}{P{2.6cm}P{1.4cm}P{1.4cm}P{2cm}P{1.9cm}P{1.9cm}}
    \toprule  
     Dataset & \# Users & \# Items & \# Interactions & \# User feature & \# Item features \\
     \midrule
     Foursquare &  2,060 & 2,876 & 27,149 & 2,108 & 347 \\
     Movielens-1M &  6,040 & 3,704 & 1,000,209  & 21 & 18 \\
     Epinions &  5,598 & 4,064 & 542,741 & 10 & 10 \\
     \bottomrule
\label{tab:data_stats}
\end{tabular}
}
\end{table}

\begin{figure*}[htp]
  \centering
  \includegraphics[height=8.5cm]{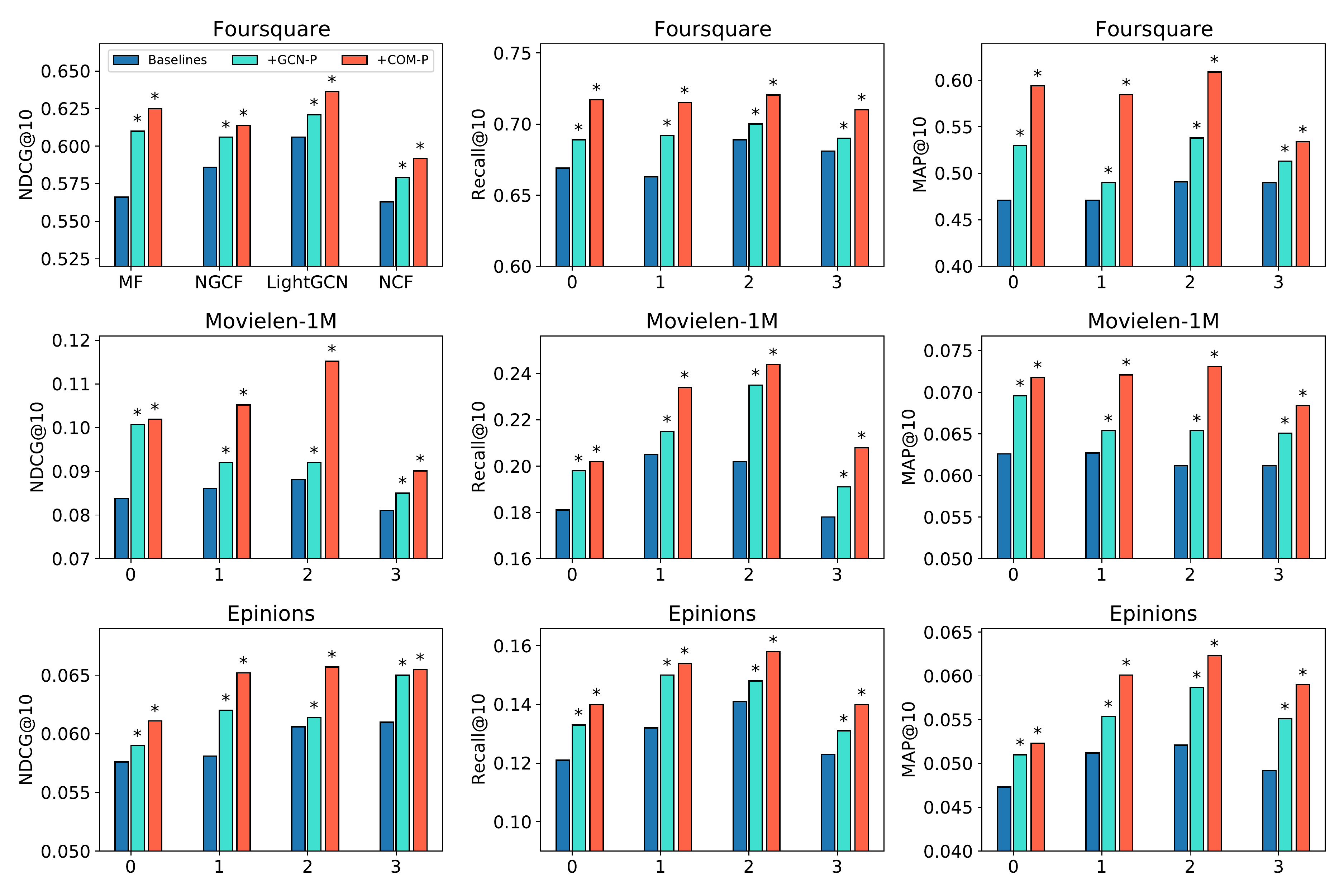}  
  \caption{\label{fig:pre-train}\iadh{Performances} comparison of the four selected existing recommenders with our \gcn{} and \com{} pre-training processes. \siwei{We use $*$ to denote a significant difference between the performances of \iadh{the} baselines and \iadh{their} pre-trained variants, according to the paired t-test \sws{with the Holm-Bonferroni correction} for p<0.01. }}
\end{figure*}

\subsection{\iadh{Experimental} Settings}
We use the leave-one-out splitting~\cite{he2017neural,rendle2009bpr,rendle2020neural} method to split the interactions of each dataset into training, validation and testing sets. 
\zms{To speed up the evaluation, we adopt the sampled metrics~\cite{he2017neural,NGCF19,rendle2020neural}, \io{which} randomly sample a small set of the non-interactive \craig{items} as negative items (rather than take all the non-interactive items as negatives) of the validation and testing sets, and evaluate the metric performance on this smaller set.}
Here, we sample 100 negative items for each user in both the testing sets and validation sets for evaluation~\cite{he2017neural,kang2018self,rendle2020neural}. \craig{However}, different from \io{prior} works~\cite{he2017neural,rendle2020neural} that only use one oracle testing set per dataset with the sampled negative items, we construct 10 different testing sets with different sampled negative items for each \iadh{dataset} with different random seeds, \iadh{in order} to reduce the evaluation bias on some specific testing \iadh{sets}. \io{Hence,} the \iadh{reported} performance of each run \iadh{is} based on the average of \iadh{the} 10 testing sets. Three ranking evaluation metrics, namely the Normalised Discounted Cumulative Gain (NDCG), Recall and Mean Average Precision (MAP) metrics, are applied for evaluating the \iadh{performances} of our \iadh{evaluated} models. 

\thispagestyle{empty}
 We evaluate the effectiveness of our pre-training models by comparing \iadh{them} with seven \iadh{existing} state-of-the-art recommendation models, where four are general \craig{(representation-based)} recommendation models, and three are recommendation models that leverage the side information of both users and items.

% Afterwards, we compare \iadh{the effectiveness of} our pre-trained models in comparison with other baselines including the HIRE~\cite{liu2019recommender}, cVAE~\cite{chen2018collective} and SSLIM~\cite{ning2012sparse} models, which also have the ability to leverage side information. To answer \textbf{RQ3}, we report the variances of the performances for our pre-trained models and their corresponding baselines where each model is run 50 times \iadh{for} 50 different randomly selected seeds. \siwei{For the \textbf{RQ4}, we vary the latent dimension to examine its effect for the pre-trained models and the baseline. We also plot the performances across different cut-off values regarding the NDCG metric to examine the performance  }
\setlist[itemize]{leftmargin=*}

\begin{itemize}
    \item \textbf{MF}~\cite{rendle2020neural}: This is the conventional matrix factorization model, which can be optimized by the Bayesian personalized ranking (BPR~\cite{rendle2009bpr}) or the BCE losses. The regularization includes the user bias, the item bias and the global bias.
    \item \textbf{NCF}~\cite{he2017neural}: This is a neural recommendation method, \craig{which learns \iadh{the} user and item embeddings while integrating \iadh{the} GMF \& MLP models to capture their non-linear feature interactions.}
    \item \textbf{NGCF}~\cite{NGCF19}: NGCF is devised to employ a multi-layer \craig{GCN} on the user-item interaction graph to propagate the collaborative signal across multi-hops user-item neighbourhoods. 
    \item \textbf{LightGCN}~\cite{he2020lightgcn}: Building on NGCF, LightGCN has fewer redundant neural components compared with the NGCF model, which makes it more efficient and effective. 
    \item \textbf{HIRE}~\cite{liu2019recommender}: \craig{This is} a recently proposed side information-aware \io{recommendation} model, which combines the flat and hierarchical side information to alleviate the challenges brought \iadh{by} the heterogeneity \iadh{of the side information}.
    \item \textbf{cVAE}~\cite{chen2018collective}: \zms{cVAE is a side information-aware recommendation model that uses the variational auto-encoder to encode the entity side information into entities for enhancing the performance.}
    \item \textbf{SSLIM}~\cite{ning2012sparse}: \craig{This is} a \io{classical} sparse linear \io{recommender}, which can \iadh{use} both the \io{users} and \io{items'} side information. In particular, we choose the binary representation to remain consistent with our feature \iadh{representations}. 
    %from their proposed side information representation methods, 
\end{itemize}

 \siwei{To demonstrate the effectiveness of our proposed pre-training \zmrs{scheme}, we apply our two pre-training models on the four \craig{existing} \ya{widely used} representation-based baselines, namely MF, NCF, NGCF and the LightGCN. We use \textbf{+GCN-P} and \textbf{+COM-P} to denote \iadh{the} two pre-training models, respectively\footnote{\iadh{E.g.} \ MF$_{+\com{}}$ stands for a model, pre-trained by Composition-GCN and fine-tuned with MF)}.}

% \todo[iadh]{What about RQ4?}

 \zms{We adopt the Adam~\cite{Adam} optimizer in \swss{both our two} pre-training models and the four fine-tuning baselines.} To determine \siwei{the values \iadh{of} all} \zmss{hyperparameter}s, we randomly sample one interaction for each user as the validation set and tune \iadh{the} \zmss{hyperparameter}s on it for all \io{of} the models. \zms{In particular, we tune our pre-training models by varying the learning rate in ~$\left \{ 10^{-2},10^{-3},10^{-4} \right \}$ and the regularization weight $\lambda$ in $\left \{ 10^{-2},...,10^{-5} \right \}$.} \zmss{The learning rates of the baseline models are also tuned according to the suggested ranges from the original papers.} The depths for all GNNs of the graph-based recommenders (i.e. NGCF and LightGCN) and the pre-training scheme are kept \iadh{to} 3 with each layer having a size of 64, while the dropout ratios of all GNNs vary among $\left \{ 0.3,0.4,...,0.8 \right \}$ as suggested in the existing literature~\cite{he2020lightgcn}. \siwei{We \craig{set} the maximum number of training epochs \iadh{to} \sws{500}; \iadh{the} batch size \iadh{to} 1000 and \iadh{the} latent dimension \iadh{to} 64 for all models. } 
 %\inote{should we justify this choice based on the literature?}}
 % Zaiqao: The batch size is not so sensitive to the performance, in literature they normally use 512 or 1024. 64 is just for a fair comparison.
 \sws{Moreover, we use \cm{an} early stopping strategy, i.e., \io{we apply a} premature stopping if \io{NDCG}@10 on the validation data does not increase for 20 successive epochs.} Note that the embedding dimension $d$ is a hyperparameter for both the pre-training models (i.e.\ the \gcn{} and \com{} models) and the fine-tuning models (i.e. the baseline models). For \iadh{a} fair comparison, we set this hyperparameter to 64, since most of our experimental models can almost \iadh{achieve} their best performances \craig{for this dimension size} across our three datasets. %\swss{\cm{We also} tune all baselines within\inote{for, or using?} the same set of hyperparameters, such as learning rate and regularization as our proposed models.} % zaiqiao: Repeated 
 A detailed analysis of the \iadh{models' performances} over \iadh{different} embedding dimensions can be found in \S\ref{sec:hyper}. For our \com{} model, the number of learnable basis vectors $b$, i.e. a \zmss{hyperparameter of the Composition-GCN~\cite{vashishth2019composition} layers, is set to 10, which is empirically tuned from \io{the} set \{5,7,10,20\} by using the validation set for \sws{all} datasets}\footnote{Note that a \swss{more} \craig{thorough} tuning of this parameter may further improve the recommendation performances, but \io{we did not observe a} clear performance trend over different $b$ values \io{in} our experiments, and under this setting we have already obtained \io{excellent} performances that can be used to draw our conclusions.}.  

%\todo[craig]{what I want to know is if your training regime is fair to the baselines - did they get the same chances for tuning of learning rates etc?}
% We consider the MF~\cite{rendle2020neural} model and different competitive neural recommenders including NGCF~\citet{NGCF19}, LightGCN~\cite{he2020lightgcn} and NCF~\cite{he2017neural}. Besides, we compare our models with the HIRE~\cite{liu2019recommender}, cVAE~\cite{chen2018collective} and SSLIM~\cite{ning2012sparse}, which have the ability to integrate the side information. 
% In particular, we compare our pre-trained models with all baselines in terms of Normalised Discounted Cumulative Gain@10 (NDCG), Recall@10 and Mean Average Precision@10 (MAP).
% All models are implemented in PyTorch using the Beta-Recsys open sourse framework~\cite{meng2020beta}. 

% \todo[iadh]{We need to ensure reproducibility by giving a full account of the experimental setting used; currently it looks like it is missing a number of details; e,g, how many epochs? do you use an optimizer, etc}
\section{Results and Analysis}\label{sec:result}
\thispagestyle{empty}
% \begin{table*}
% \small
% \centering
% \begin{tabular}{P{2.1cm}P{1.3cm}P{1.3cm}P{1.3cm}P{1.3cm}P{1.3cm}P{1.3cm}P{1.3cm}P{1.3cm}P{1.3cm}P{1.3cm}P{1.3cm}P{1.3cm}}
%      \multicolumn{5}{c}{Foursquare}\multicolumn{4}{c}{Movielens\_100k}\multicolumn{4}{c}{Movielens-1M}\\
%      Model & NDCG@10 &Precision@10 & Recall@10 &MAP@10\\
%      MF$_{+\com{}}$ & $\boldsymbol{0.621}$ & $\boldsymbol{0.0706}$&$\boldsymbol{0.706}$&$\boldsymbol{0.593}$ \\
%      MF$_{+\gcn{}}$ &0.571 & 0.0682 & 0.689 & 0.530 \\
%      MF &0.516 & 0.0661 & 0.669 & 0.471\\
%      \hline
%      \multicolumn{5}{c}{Movielens-1M}\\
%      Model & NDCG@10 &Precision@10 & Recall@10 &MAP@10\\
%      MF$_{+\com{}}$ & 0.0825& 0.0180& 0.180& 0.0535 \\
%      MF$_{+\gcn{}}$ &0.0804& 0.0178& 0.178& 0.0517 \\
%      MF &0.0738 &0.0171& 0.171& 0.0451\\
%      \hline
%      \multicolumn{5}{c}{Movielens\_100k}\\
%      Model & NDCG@10 &Precision@10 & Recall@10 &MAP@10\\
%      MF$_{+\com{}}$ & 0.0825& 0.0180& 0.180& 0.0535 \\
%      MF$_{+\gcn{}}$ &0.0804& 0.0178& 0.178& 0.0517 \\
%      MF &0.0738& 0.0171& 0.171& 0.0451\\
%      \hline
% \end{tabular}
% \caption{MF model results}
% \label{tab:stat_all}
% \end{table*}

In this section, we report the \io{results} \io{obtained from} five main experiments aimed at answering the research questions listed in \S\ref{sec:rqs}. In particular, we first address \textbf{RQ1} by analysing whether our two pre-training models can help to improve the four existing \iadh{representative} recommenders. Then, we further compare the performances of these models with three state-of-the-art recommenders \iadh{that leverage} both \io{the} \iadh{users' and items' side information} (\textbf{RQ2}). \zms{We provide an ablation study \swss{where we randomly drop \{20\%,40\%,60\%,80\%\} of \io{the} entity features during the pre-training process} \io{in order to seek an} answer \io{to} \textbf{RQ3}.} To answer \textbf{RQ4}, we conduct experiments over different random seeds, and \iadh{analyze} the standard deviations of these models. Finally, \io{to answer \textbf{RQ5}}, we \iadh{provide} a detailed analysis of the \iadh{performances} of our pre-training models on different embedding \iadh{dimensions} and different \siwei{cut-offs for \iadh{the} recommended items}.

\subsection{Effectiveness of Pre-training}
\label{sec:effect_pre-train}

To validate \iadh{the} effectiveness of our pre-training models, we compare the performances of the four selected recommender models (i.e.\ MF~\cite{rendle2020neural},  NCF~\cite{he2017neural}, NGCF~\cite{NGCF19} and LightGCN~\cite{he2020lightgcn})
 \craig{with} their \craig{pre}-training variants under the \io{pre-training} processes \io{defined by} our two proposed \siwei{pre-training approaches} (i.e. \gcn{} and \com{}).  \fig{}~\ref{fig:pre-train} reports the recommendation \iadh{performances} comparison \iadh{in terms of} the NDCG, Recall and MAP \iadh{metrics} at \siwei{a \craig{rank} cut-off of 10} -- \siwei{later in \S\ref{sec:hyper}, we use different \iadh{cut-off} values for \iadh{a} further detailed analysis.} \craig{In the figure, \io{performance improvements} that significantly  (according to a \sws{paired t-test with the Holm-Bonferroni correction}) outperform the corresponding baseline recommender model are denoted with * ($p<0.01$).}
%  We only report $k=10$ here as the performances \iadh{using} other $k$ values resulted in the same \iadh{trends and} conclusions, and the \iadh{actual} performances \iadh{using} different $k$s \iadh{values} are reported in \iadh{Section~}\ref{sec:hyper}.
From \fig{} \ref{fig:pre-train}, we can \iadh{clearly observe} that, over the three \iadh{used} datasets, all the selected \io{4} recommender models \craig{exhibit significantly improved performances} when our pre-training \siwei{approaches} are applied.
Moreover, we can also see that \siwei{a baseline pre-trained with our \com{} approach can always outperform the baseline pre-trained with the \gcn{} approach.}
% with our \com{} pre-training model consistently outperform the corresponding models with our \gcn{} pre-training model. 
This is somewhat expected, as the \com{} approach \iadh{is} trained using the multi-relational graphs constructed from \iadh{the} \io{entities'} side information. \siwei{Therefore, the \com{} approach is capable of capturing the heterogeneous \iadh{relations} between entities within the side information, in contrast to the \gcn{} approach, which can only leverage the similarity between each feature vector.}
%\inote{ambiguous, who which corresponds to? Split sentence} 
% able to capture the heterogeneous relations between entities within their side information.
% In particular, we observe between \siwei{8.4\%} to \siwei{37.2\%} \iadh{improvements} over the four selected models in the Foursquare dataset under the NDCG metric and \siwei{2.74\%} to \siwei{23.2\%} \iadh{improvements} under the Recall metric.
 
To conclude on \textbf{RQ1}, we have shown that \siwei{our \io{two proposed} pre-training approaches can effectively leverage various \io{users} and items' side information, thereby enhancing \iadh{the} existing representation-based baseline recommender models \zmrs{with significant performance improvements}, consistent across the three \io{used} datasets, three measures and four baselines.}

%  our pre-training models indeed help \iadh{to enhance the performances of} the existing recommendation systems, and \iadh{that} our \com{} pre-training model is more effective in helping the existing recommendations systems \iadh{better} leverage the \iadh{various available} side information.

%  \todo[iadh]{Is the comparison fair? the enhancements might be coming from the side information, not from the pre-training process itself.}
 \thispagestyle{empty}

 \subsection{Effectiveness of Integrating Side Information}

\begin{table}
\caption{\label{tab:all_side_models}\iadh{Performances} comparison of recommenders with side information. The best and second best performances are marked in boldface \io{or} underlined, respectively. \sws{We use $*$ to denote a significant difference between the performances of the side information-aware baselines and the best proposed model, according to the paired t-test with the Holm-Bonferroni correction for $p<0.01$.} }
\label{tab:side}
\small
\centering
\resizebox{78mm}{!}{
\begin{tabular}{P{2.0cm}P{0.7cm}P{0.7cm}P{0.7cm}P{0.7cm}P{0.7cm}P{0.7cm}}
     \toprule
     \multirow{2}{*}{Model} &\multicolumn{2}{c}{Foursquare} & \multicolumn{2}{c}{Movielens-1M}& \multicolumn{2}{c}{\sws{Epinions}}\\
      & NDCG  & MAP & NDCG & MAP & NDCG & MAP\\
     \midrule
     HIRE & 0.5232$^{*}$&  0.4575$^{*}$& 0.0997$^{*}$ & \underline{0.0692}$^{*}$ & 0.0667$^{*}$  &0.0518$^{*}$\\
     cVAE &0.5326$^{*}$&  0.4438$^{*}$& 0.0678$^{*}$ & 0.0453$^{*}$ & 0.0532$^{*}$ &0.0410$^{*}$\\
     SSLIM & 0.5894$^{*}$ & 0.4691$^{*}$& 0.0655$^{*}$ & 0.0441$^{*}$& 0.0533$^{*}$& 0.0401$^{*}$\\
     MF$_{+\gcn{}}$ &0.6206&	0.5901 & 0.0979& 0.0646 & 0.0527&0.0434\\
     MF$_{+\com{}}$ & \underline{0.6249} & \underline{0.5944} &  \underline{0.1019} & \textbf{0.0718} & 0.0587 & 0.0498\\
     NGCF$_{+\gcn{}}$ &0.6016& 0.5683 &0.0713& 0.0450 & 0.0708& 0.0454\\
     NGCF$_{+\com{}}$ & 0.6138& 0.5844& 0.0752& 0.0461 & \underline{0.0719} &0.0485 \\
     LightGCN$_{+\gcn{}}$ & 0.6162& 0.5940& 0.0952& 0.0631 & 0.0717& \underline{0.0594}\\
     LightGCN$_{+\com{}}$ &\textbf{0.6364}& \textbf{0.6089}& \textbf{0.1068}& 0.0689 & \textbf{0.0792}& \textbf{0.0623}\\
     NCF$_{+\gcn{}}$ &0.5677& 0.4939& 0.0870& 0.0551 & 0.0620& 0.0531\\
     NCF$_{+\com{}}$ &0.6021 &0.5340 & 0.0913 & 0.0584 & 0.0691 & 0.0583\\
     \bottomrule
\end{tabular}}
\end{table}

\begin{table}
\caption{Standard deviations (denoted std.) and means of the NDCG@10 performances over 50 random seeds. Lower standard deviation (in bold) means \iadh{a} better stability. }
% \inote{where is movielens}
\label{tab:stability}
\resizebox{63mm}{!}{
\begin{tabular}{P{2.1cm}P{0.7cm}P{0.7cm}P{0.7cm}P{0.7cm}}
\toprule
     \multirow{2}{*}{Model} &\multicolumn{2}{c}{Foursquare} &  \multicolumn{2}{c}{\sws{Epinions}}\\
      & std.  & mean & std. & mean\\
     \midrule
     MF & 0.0197&0.5163& 0.0127&0.0501\\
     MF$_{+\gcn{}}$ &0.0046& 0.6206 & 0.0049&0.0527  \\
     MF$_{+\com{}}$ &\textbf{0.0029} &0.6249 & \textbf{0.0043} & 0.0587\\
     \hline
     NGCF & 0.0237& 0.4162& 0.0108&0.0681\\
     NGCF$_{+\gcn{}}$ & 0.0053&0.6016  & 0.0042& 0.0708 \\
     NGCF$_{+\com{}}$ & \textbf{0.0044}& 0.6138&\textbf{0.0038} &  0.0719\\
    \hline
     LightGCN & 0.0209&0.4365& 0.0092&0.0706\\
     LightGCN$_{+\gcn{}}$ & 0.0050& 0.6262& 0.0039& 0.0717 \\
     LightGCN$_{+\com{}}$ & \textbf{0.0039}& 0.6264& \textbf{0.0031} & 0.0792 \\
    \hline
     NCF & 0.0283&0.4621& 0.0159&0.0591\\
     NCF$_{+\gcn{}}$ & 0.0091&0.5777 &0.0079 & 0.0620 \\
     NCF$_{+\com{}}$ & \textbf{0.0086}&0.6021 &  \textbf{0.0064}& 0.0691 \\
     \bottomrule
\end{tabular}
}
\end{table}

Having shown that our proposed \iadh{pre-trained} approaches are effective \io{at} \iadh{enhancing the performances of} the existing recommendation systems \iadh{through the leveraging of} side information, we next examine whether these \iadh{recommendation} systems with our pre-training scheme %\inote{1 scheme, 2 models right? Siwei: Right}
\io{perform} better than the \io{existing} state-of-the-art recommendations systems, \iadh{which leverage an} integration scheme to incorporate \iadh{the} entity side information. To answer this, we further compare \swss{our pre-training models} 
%\inote{vague, which ones? the proposed pre-training models or the ones from the literature?} 
with three state-of-the-art recommenders \iadh{where} both side information of users and items are \iadh{used}. 
% \footnote{\craig{Due to the increased experimental complexity \iadh{for} the larger dataset, the performances reported in \tab{} \ref{tab:side} for both Foursquare and \sws{Epinions} are the means of 50 runs, while} \iadh{for} Movielens-1M, \iadh{they are} the means of 10 runs, all with different random seeds.}
\tab{} \ref{tab:side} reports the recommendation \iadh{performances in terms of the} NDCG and MAP  metrics at rank \craig{cut-off} 10 for each model across all three datasets. From \io{\tab{}~\ref{tab:side}}, we \io{observe} that although the \craig{relative performance ranking of systems} \iadh{is} different across \iadh{the used} datasets, the three \iadh{baselines} (i.e.\ HIRE, cVAE and SSLIM) \craig{do not} \iadh{achieve} the \craig{highest} \iadh{performances on any of the used} datasets. \craig{Specifically, the \sws{LightGCN}$_{+\com{}}$ model performs the best \iadh{on both the} Epinions and \sws{Foursquare} datasets, \io{outperforming} the NDCG@10 of the HIRE side information-aware baseline by \sws{18.7}\% (\sws{0.0667} $\rightarrow$ \sws{0.0792}) and 21.6\% (0.5232 $\rightarrow$ 0.6364), \io{respectively}. For the \swss{Movielens-1M} dataset, MF$_{+\com{}}$ achieves the best performance \io{on} the MAP metric, outperforming the HIRE model by 3.76\% (0.0692 $\rightarrow$ 0.0718). }

\craig{Overall,} \io{in} answer \io{to} \textbf{RQ2}, we have shown that the four selected \io{representative recommendation} models with our \com{} pre-training scheme can outperform the other three baseline models that leverage entity side information on all the three \io{used} \craig{datasets}, while with our \zms{\gcn{}} pre-training scheme they can still outperform the other three baseline models in most cases on the Foursquare and \sws{Epinions} datasets.

\subsection{Ablation Study of Side Information}

\begin{figure*}[htp]
    \includegraphics[width=0.75\textwidth]{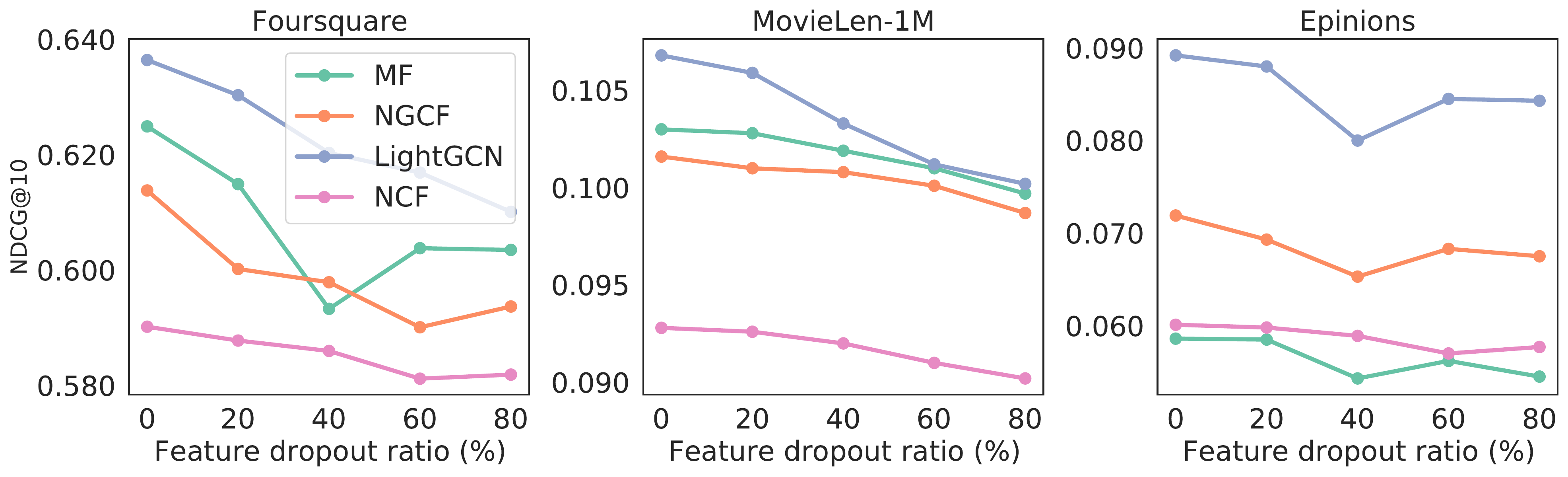}
    \caption{Results of the baselines deploying the \com{} model over different dropout ratio of entity features.}
    \label{fig:ablation}
\end{figure*}

\zms{Having \io{observed} that all the \iadh{evaluated} existing \swss{baseline} models are significantly improved by our pre-training models, we \io{now check} whether these improvements are actually \io{the result of using} the within-entity knowledge that is pre-trained by our models \sws{(\textbf{RQ3})}. To answer this \io{research question}, we conduct an ablation study \swss{to examine the effect of randomly removing entity features, \io{thereby} revealing the connection between the performance improvements and the within-entity knowledge.}
%\inote{unclear why this would answer RQ3, we need a sentence that explains the used methodology}.
Specifically, we randomly drop different proportions (\{20\%, 40\%, 60\%, 80\%\}) of entity features during the pre-training process, and evaluate the recommendation \io{performances} of the \io{fine-tuned} models under these pre-trained embeddings.} \zms{ \io{Figure~\ref{fig:ablation} reports the obtained results}. Here, \io{for space reasons}, we only report the \io{results of} \com{} since \io{we observed the same trend with} \gcn{}. From Figure~\ref{fig:ablation}, we \io{can see} \craig{that} all the NDCG@10 performances of the four \swss{fine-tuned} models decrease as the \io{features} dropout ratio increases from 0\% (i.e. no \swss{dropped features}) to 80\% (\swss{80\% of features \io{are} dropped}) in all the three datasets. \io{This \zmss{result} suggests that} randomly dropping entity features \io{does hurt} the overall recommendation performance. This result also \io{suggests} %\inote{unclear how unless you explain your methodology/rationale for using this experiment to answer RQ3} 
that the performance improvements are indeed gained from the entity features and our pre-training models are able to \swss{accurately} capture the within-entity knowledge from \io{the} \swss{entities'} side information.}
%From Figure~\ref{fig:ablation}, we find that randomly dropping entity features will hurt the overall recommendation performance, hence the NDCG@10 performance of all models decreases when feature drop ratio increases from 0 i.e. when no feature is dropped. As a result, this ablation study generally helps us to demonstrate that these performance improvements do come from the pre-training using the entity features. However, it's also noticeable that different with the case of Movielen 1M dataset when constant decreases are observed for all models, for the other two datasets, performance sometimes rebound as the feature drop ratio increases. This is also reasonable because for those two datasets, we use bag-of-words selected from reviews as their feature vectors, which are less reliable than features provided in the Movielen dataset such as ''occupation'' for users and ''genre'' for movies. 

% \zmsc{update ablation study results.}

\thispagestyle{empty}

\subsection{Stability Analysis}
\label{sec:stab}
%\ianote{I would not use this adjective; as a reviewer will reject saying approach is too simple!}
\io{We have demonstrated in the previous sections the general applicability and effectiveness of our proposed pre-training scheme.}
%From the previous \iadh{results} and analysis, \swss{we have seen} the general applicability of our proposed pre-training scheme \swss{and} we obtain strong empirical performance improvements for each baseline \inote{this sentence is weird; "despite" suggests there was a bad result, but so far, all seems good, so i'm confued what this sentence is for?}. 
To \craig{address} \textbf{RQ4}, we calculate the standard deviations of \craig{the NDCG@10 performances of} the baseline models (i.e.\ MF, NCF, NGCF and LightGCN) and their enhanced variants by our pre-training models (\io{indicated} with the subscripts $_{+\gcn{}}$ and $_{\com{}}$ for each baseline). \io{\tab{}~ \ref{tab:stability} presents the obtained results \swss{on the Foursquare and Epinions datasets\footnote{\cm{Due to its size, we exclude the Movielen-1M (it has around $\times$ 2 and $\times$ 4 more interactions than Foursquare and Epinions) as it would take an excessive time to repeat experiments 50 times.}}}}.
%the \iadh{results} of which \iadh{are} shown in \tab{} \ref{tab:stability}. 
\iadh{From the table, we} observe that our pre-training scheme \craig{markedly} \iadh{improves} \swss{the performances and stabilities} of \swss{all} the \io{used} baselines, with much smaller \swss{observed} standard deviations in each paired comparison of a baseline model with and without the use of the pre-training scheme. {Furthermore,} our \com{} pre-training model consistently outperforms the \gcn{} pre-training model, which indicates that the more complex relationships captured from the entity features by the multi-relational graphs can make the pre-training model \io{more effective in obtaining} a better embedding initialization for the recommendation \io{model} \io{and achieving a} more stable recommendation performance.

% \todo[iadh]{To complete highighting some results, providing further insights on results, etc}

%todo: update new result for the baseline models

\subsection{Hyperparameter Analysis}
\label{sec:hyper}

\begin{figure}[htp]
     \centering
     \begin{subfigure}[b]{0.23\textwidth}
         \centering
         \includegraphics[width=1\textwidth]{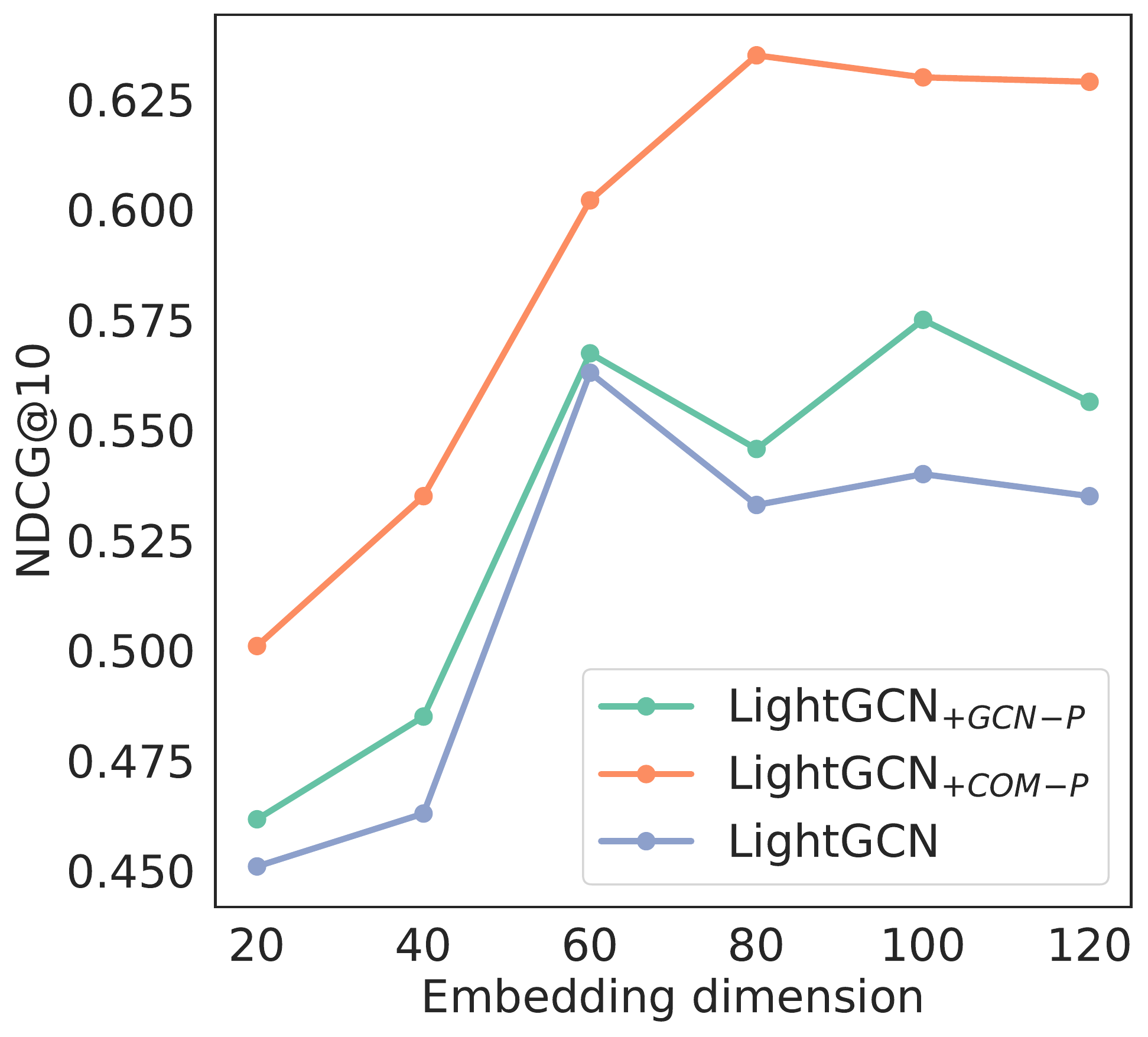}
         \caption{\footnotesize NDCG vs.\ Embedding size}
         \label{fig:dim}
     \end{subfigure}
     \begin{subfigure}[b]{0.23\textwidth}
         \centering
         \includegraphics[width=1\textwidth]{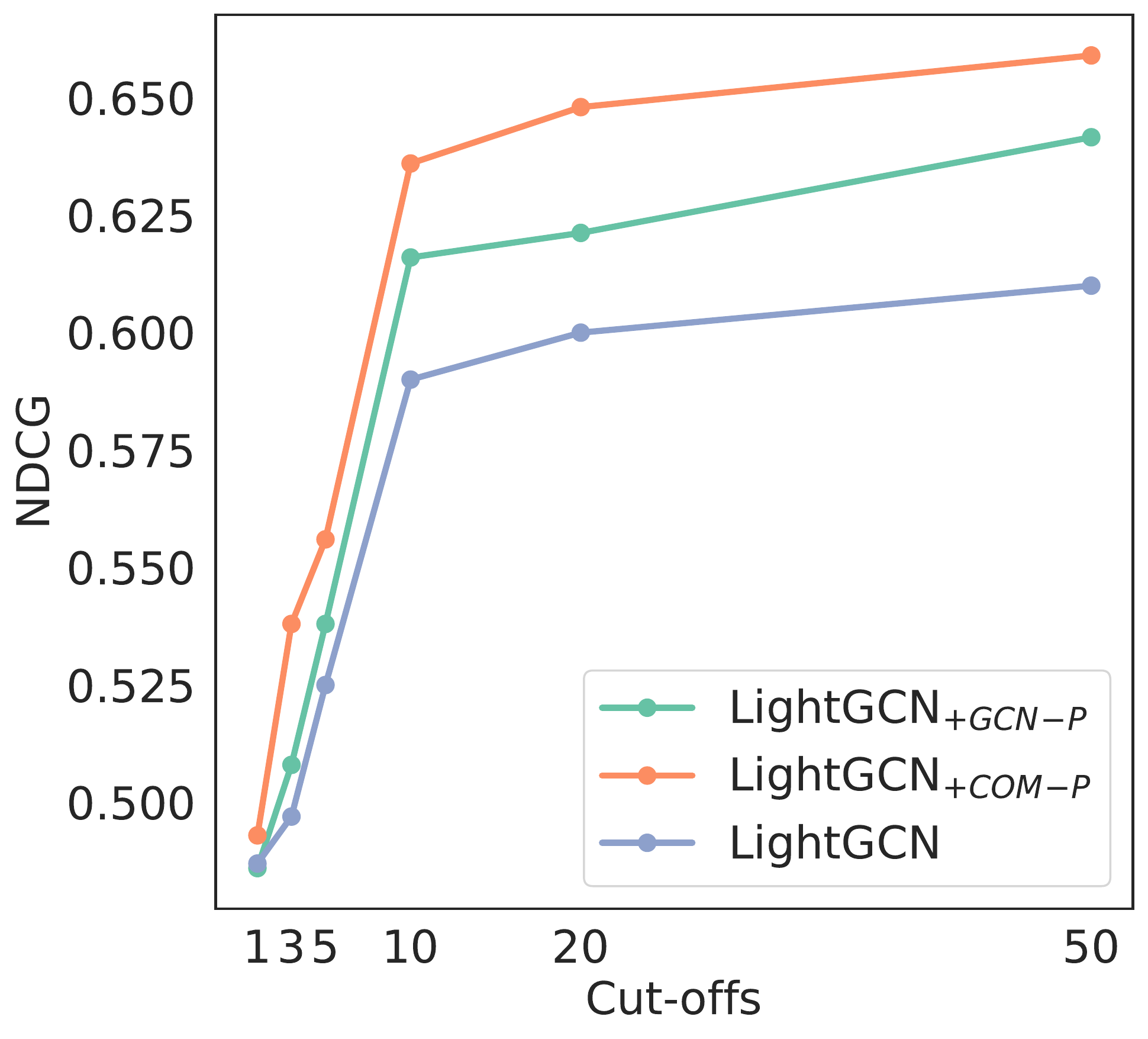}
         \caption{\footnotesize NDCG vs.\ $k$}
         \label{fig:cut}
     \end{subfigure}
     \caption{The comparison between the \swss{LightGCN} baseline with its pre-training variants on the Foursquare dataset.}
\end{figure}

% \begin{figure}
%     \includegraphics[width=0.48\textwidth]{figs/dim.pdf}
%     \caption{The comparison between the \swss{LightGCN} baseline with its pre-trained models on the Foursquare dataset..}
%     \label{fig:dim}
% \end{figure}

To answer \textbf{RQ5}, we study how the \ya{embedding} dimension affects the recommendation performance. Figure~\ref{fig:dim} shows the \iadh{comparative results} of the performances of the \swss{LightGCN} model and \zmss{\io{its} two variants} \io{with} our \zmss{pre-training} models\footnote{\zmss{For space reasons, here we only report the \io{results} for the \swss{LightGCN} model on the Foursquare dataset, since other fine-tuning models and other datasets \io{produced} the same conclusion.}}. 
%\inote{why this model in particular? what about the others?}
%\inote{why this dataset? we need footnotes to say if trends are the same with other models/datasets?}
As a sanity check, we \iadh{observe} that the size of \iadh{the embedding} dimension \iadh{\io{does} indeed affect} the final recommendation \iadh{performances} of all the three \iadh{evaluated} models.
%\inote{(NCF and its 2 variants?)}
We can also observe that when the size of \iadh{the} latent \craig{dimensions} is \cmrs{$\leq$40}, \iadh{the} \swss{LightGCN} \ya{model} and both of its pre-trained variants show relatively \iadh{poor performances}, which can be \iadh{further} boosted when the dimension \craig{size increases}; almost all models' performances reach their highest points when the dimensions are between 60 to 80. \craig{Recall} that, for a fair comparison, we \iadh{fixed} the embedding dimension \craig{to} 64 for all \iadh{the} implemented models, which can be \ya{further} justified from \iadh{these obtained results}. Moreover, Figure~\ref{fig:dim} \craig{demonstrates} that our pre-training scheme can bring consistent improvements to the exiting models across different embedding dimensions. We also plot the performances evaluated by NDCG over different cut-offs ($k$) of the \zmrs{items} ranking in Figure~\ref{fig:cut}. We can also see that our pre-training models consistently \craig{improve over the \swss{LightGCN} model for} different cut-offs. \zm{In particular, we see that even \craig{for} some \craig{low rank cut-offs}  (e.g.\ $k=\{1, 3, 5\}$) or \io{for} \craig{deep cut-offs (e.g.\ $k=50$)} our \com{} \iadh{model} can still enhance the \swss{LightGCN} model with a marked improvement, which means that our per-training \io{approach} can help improve \swss{the recommendation performance under different circumstances when different amount of items (i.e. cut-off values) are chosen to be exposed to \io{the} users.}}
%\inote{vague, is that a suggested idea?} 

% the recommendation systems by adaptively controlling the display of items.} 
%\iadh{Note that the performances evaluated by other metrics and other datasets are not reported here, since they produce the same trends and conclusions} \inote{maybe comes too late, say from the ouset, and recall finish by recalling that we observed the same trends on other models, datasets and metrics?}.

\thispagestyle{empty}
\section{Conclusions}\label{sec:concolusion}
\thispagestyle{empty}
In this paper, we \iadh{introduced} a \iadh{novel} pre-training scheme for the recommender systems to leverage the entity side information \zmrs{in a general \ya{manner}}. In particular, we \iadh{proposed} two models for pre-training the entity representations to capture the \zmss{within-entity contextual knowledge}, based on the graphs constructed from the entity side information. \swss{The extensive evaluation} of our pre-training models with the fine-tuning of four existing representation-based recommender models showed that effectively pre-training \iadh{the} embeddings with both \io{the} users and items' side information improved these existing models in terms of both effectiveness (\craig{always significantly, see Figure~\ref{fig:pre-train}) and stability (see Table~\ref{tab:stability}}). \swss{Furthermore, compared to the existing state-of-the-art recommender baselines, which integrate the same side information, our pre-training models exhibited upto 7\% improvement in NDCG@10 for the Movielen-1M dataset, 21\% improvement on the Foursquare dataset and 48\% improvement on the Epinions dataset (see Table \ref{tab:side}).} We also showed that the \com{} model, pre-trained under the multi-relational graphs, can always outperform the \gcn{} model, pre-trained under the single-relational graphs (see Figure~\ref{fig:pre-train}), which \io{suggests that} more information \io{is} captured by the multi-relational graphs. Our pre-training scheme provides a general \zmss{framework} \iadh{for leveraging} side information, \io{which} can be used to enhance any general representation-based recommendation \io{model}. \iadh{As} future work, we \io{aim} to investigate \cm{using}  other types of graph neural networks \cm{in} the pre-training \swss{scheme}. We will also explore whether \iadh{the} \ya{proposed} pre-training scheme can benefit other recommendation scenarios, such as \iadh{in} sequential and session-based recommendations.

\bibliographystyle{ACM-Reference-Format}
\clearpage
\bibliography{main}
\thispagestyle{empty}
\end{document}